%% file: Main.tex
\documentclass{jfm}
\usepackage{hyperref}
\usepackage{graphicx}
\usepackage{epstopdf, epsfig}
\usepackage{amsmath,bm}
\usepackage{subcaption}

\usepackage{siunitx}
\usepackage{booktabs}
\usepackage[ruled,vlined]{algorithm2e}
\usepackage{esint}
\usepackage{cleveref}
\usepackage{soul}
\usepackage{xcolor}
\usepackage{ltablex}
\input{Notations.tex}

\definecolor{revision}{rgb}{1, 0, 0}
\definecolor{ForestGreen}{RGB}{34,139,34}

\usepackage[commentmarkup=footnote]{changes}
\definechangesauthor[name=Lusseyran, color=orange]{FL}
\definechangesauthor[name=Noack, color=red]{BN}

\crefname{section}{§}{§§}
\Crefname{section}{§}{§§}

\shorttitle{Stabilization of the open cavity flow}
\shortauthor{G.~Y.~Cornejo Maceda, E.~Varon, F.~Lusseyran, and B.~R.\ Noack}

\title{Stabilization of a multi-frequency open cavity flow with gradient-enriched machine learning control}

\author{
Guy Y.\ Cornejo Maceda\aff{1,2},
  Eliott Varon\aff{2},
  Fran\c{c}ois Lusseyran\aff{2},
  \and Bernd R.\ Noack\aff{1,3}\corresp{\email{bernd.noack@hit.edu.cn}}
  }

\affiliation{
\aff{1} School of Mechanical Engineering and Automation,
Harbin Institute of Technology (Shenzhen), 
University Town, Xili,
Shenzhen 518055, People's Republic of China.
\aff{2} Universit\'{e} Paris-Saclay, CNRS, Laboratoire Interdisciplinaire des Sciences du Num\'{e}rique, 91400, Orsay, France.
\aff{3} Hermann-F\"ottinger-Institut,
Technische Universit\"{a}t Berlin, M\"{u}ller-Breslau-Stra{\ss}e 8, D-10623 Berlin, Germany.
}

\begin{document}
\maketitle

\input{S0} 
\input{S1} 
\input{S2} 
\input{S3} 
\input{S4} 
\input{S5} 
\input{S6} 
\input{Acknowledgements}

\appendix
\input{SA} 
\input{SB} 
\input{SC} 
\input{SD} 
 
\bibliographystyle{jfm}
\bibliography{Main,Main_Bernd,BIB_Cavity}

\end{document}

%% file: Notations.tex
\newcommand{\KgMLCRa}{K^{\textrm{I}}} 
\newcommand{\JgMLCRa}{J^{\textrm{I}}} 
\newcommand{\JagMLCRa}{{J_a}^{\textrm{I}}} 
\newcommand{\JbgMLCRa}{{J_b}^{\textrm{I}}} 
\newcommand{\siggMLCRa}{{\widetilde{\sigma}}^{\textrm{I}}} 
\newcommand{\ARa}{A^{\textrm{I}}} 
\newcommand{\KgMLCRb}{K^{\textrm{II}}} 
\newcommand{\JgMLCRb}{J^{\textrm{II}}} 
\newcommand{\JagMLCRb}{{J_a}^{\textrm{II}}} 
\newcommand{\JbgMLCRb}{{J_b}^{\textrm{II}}} 
\newcommand{\siggMLCRb}{{\widetilde{\sigma}}^{\textrm{II}}} 
\newcommand{\ARb}{A^{\textrm{II}}} 

\newcommand{\Nvar}{N_{\rm vr}} 
\newcommand{\Ncst}{N_{\rm cr}} 
\newcommand{\Ninstrmax}{N_{\rm inst, max}} 
\newcommand{\Nb}{N_{\rm b}} 
\newcommand{\Pc}{P_c} 
\newcommand{\Pm}{P_m} 
\newcommand{\Nmc}{N_{\rm MCS}} 
\newcommand{\Nsub}{N_{\rm sub}} 
\newcommand{\Nig}{N_{\rm p}} 

\newcommand{\Str}{St} 

%% file: S0.tex
\begin{abstract}
We stabilize an open cavity flow experiment to 1\% of its original fluctuation level.
For the first time, a multi-modal feedback control is automatically learned for this configuration.
The key enabler is automatic in-situ optimization of control laws
with machine learning augmented by a gradient descent algorithm,
named gradient-enriched machine learning control \citep[gMLC]{CornejoMaceda2021jfm}.
gMLC is shown to learn one order of magnitude faster than MLC \cite[MLC]{Duriez2017book}.
The physical interpretation of the feedback mechanism 
is assisted by  a novel cluster-based control law visualization 
for flow dynamics and corresponding actuation commands.
Starting point of the control experiment are two unforced open cavity benchmark configurations: 
a narrow-bandwidth regime with a single dominant frequency 
and a mode-switching regime where two frequencies compete.
The feedback control commands the DBD actuator located at the leading edge.
The flow is monitored by a downstream hot-wire sensor over the trailing edge.
The feedback law is optimized with respect to the monitored fluctuation level.
As reference,  the self-oscillations of the mixing layer are mitigated with steady actuation.
Then, a feedback controller is optimized with gMLC.
As expected, feedback control outperforms steady actuation by achieving both, 
a better amplitude reduction and a significantly smaller actuation power,
about 1\% of the actuation energy required for similarly effective steady forcing.
Intriguingly, optimized laws learned for one regime performs well for the other untested regime as well.
 The proposed control strategy can be expected to be applicable for many other shear flow experiments. 
\end{abstract}

%% file: S1.tex
\section{Introduction}\label{sec:introduction}

Open cavity oscillations occur in many ground and airborne transport vehicles, 
like wheel casings or bogeys,
and significantly contribute to aerodynamic drag and noise.
Active model-based control has been applied with large success 
to the stabilization of these oscillations \citep{Rowley2006arfm,Sipp2010amr}.
In this study, we aim at fast self-learning feedback which simplifies the development of control
and extends the applicability to nonlinear dynamics.
Encouraged by results for wake stabilization \citep{CornejoMaceda2021jfm},
 we apply gradient-enriched machine learning control to an experiment.

Open cavity flows typically feature mono-mode and multi-frequency regimes 
depending on the configuration.
The oscillatory dynamics gathers most of the mechanisms responsible for nonlinear turbulence interactions.
Yet, the self-organization of the spatial structures is still highly coherent
and driven by global instability \citep{Huerre1998}.
Our configuration has a moderate Reynolds number ($\Rey_L\approx10^4$).
The length-to-depth ratio is around 1.7 and thus between a shallow and deep cavity.
With increasing incoming velocity
an open cavity  successively features, first an intra-cavitary centrifugal instability 
then self-sustained oscillation of the mixing layer \citep{Rowley2006arfm,Basley2014,feger2019cfm}.
The dynamics of the interaction between an incoming boundary layer and a rectangular cavity depends on six parameters: 
the ratios of the three spatial dimensions of the cavity 
(in particular the length $L$ and depth $D$ of the cavity), 
the momentum boundary layer thickness $\theta_0$ at the upstream edge, 
the incoming velocity $U_\infty$ 
and the Mach number for compressible flows.
By focussing on the two main characteristic numbers, 
$L/D$ and $\Rey_L=U_\infty L / \nu$ 
it is possible to scan a wide range of dynamics, 
from a single mode regime to spectra with rich dynamics including coupled modes \citep{Kegerise2004pof}.
This, in addition to the practical implications, is the reason for the repeated interest in this flow pattern from pioneering work \citep{Rossiter1964,gharib_roshko_1987} to the present day. 

Current studies of the cavity focus on a wide range of industry  applications.
In the transport field, due to engineering and manufacturing constraints, 
most of ground and airborne transport vehicles 
include cavities, e.g., wheel casings and bogeys,
whose interaction with low or high-speed flows 
is responsible for parasitic drag and flow-induced noise.
For German high-speed trains, 
the underbody with cavities account for 61\% 
of the aerodynamic drag and the gaps between the wagons for another 5\% \citep{Hucho2002book}.
At high-speeds such as $\SI{300}{\kilo\meter\per\hour}$, 
noise is increased by more than $\SI{14}{\decibel}$ due to cavity fluctuations \citep{Wang2014nvc}.
Landing gear bays on passenger airplanes produce strong noise and represents up to 30\% of the total noise
\citep{LiBo2020pf}.
For low-speed transports such as cars, the airflow can excite flow oscillations 
in the cavity to form resonance and noise sources, resulting in body resistance and noise nuisances for the passengers \citep{Kook1997ncej}.
Hence,  cavity flow  control is of large engineering interest.

The control of the cavity relies on the mitigation of the mixing layer 
by suppressing the feedback mechanism between the vortex formation and the impinging vortex recirculation flow.
The control can be achieved in a passive manner by modifying the geometry of the configuration or in an active manner by injection energy to the flow.
Passive devices for control include fences, spoilers, ramps, cylinders, rods \citep{Stanek2003asme,Ukeiley2004aiaa,Keirsbulck2008cjp,Panickar2008asme,ElHassan2017cjp}.
Modifications of the cavity leading edge affects the shear-layer formation \citep{Ahuja1995nasa} 
and also the trailing edge 
to reduce the sound wave generation 
at the impinging point \citep{Pereira1994aiaa}.
Porous walls have also been employed to reduce the feedback excitation near the leading and trailing edge \citep{Wilcox1988aiaa,Stallings1994citeseer}.
However, most passive devices imply parasitic drag during cruise.

On the other hand, active control may improve performance with low intrusion into the flow, 
a large frequency bandwidth and the ability to adapt to the flow response.
Noteworthy examples of model-free open-loop control 
include the stabilization of the laminar flow 
with high-frequency forcing \citep{Sipp2012jfm,Kreth2020aiaa} and pressure fluctuations mitigation for supersonic flows based on resolvent analysis \citep{Liu2021jfm}.
In contrast, most closed-loop control relies on models as a simple representation of the dynamics.
For instance, \citet{Barbagallo2009jfm} develop
a Galerkin model with global modes of the flow that preserve the input-output behavior. 
As further example, \citet{Nagarajan2018cf} achieved noise reduction with a reduced-order model including the control effect.
For quasi-periodic dynamics, an iterative method for weakly nonlinear model 
was able to completely stabilize the flow \citep{Leclercq2019jfm}.

Feedback  controllers based on linear models have also been successfully employed 
to mitigate the oscillations of the flow \citep{Illingworth2012jfm} and noise suppression \citep{Rowley2006jfm}.
Finally, we note one of the very first and remarkable closed-loop control studies by \citet{gharib1985effect} on an open cavity in a water canal.
We refer to \citep{Cattafesta2008pas} for a review on past successes of active flow control on the cavity.
A well-known effect of linear control is the shift of the oscillations of the cavity to other Rossiter modes \citep{Cabell2002ac,Williams2000asme} resulting in multi-frequency regimes.
Mode-switching regimes present a challenge for control design 
as it needs to include large bandwidths and an adequate time response \citep{Samimy2007jfm}.
Linear closed-loop control on an experimental cavity for multi-frequency control has been achieved by augmenting  the controller with well-placed zeros \citep{Yan2006aiaa}.
\citet{Samimy2007jfm} manage to control multiple frequencies by incorporating several models in linear quadratic optimal controllers.

Building a control-oriented model is often limited due to the nonlinearities of the flow 
including frequency crosstalk and time delays between the actuation and sensing.
Therefore, we choose model-free approaches based on machine learning to achieve multi-modal control.
Machine learning control \citep[MLC]{Duriez2017book} based on genetic programming \citep{Dracopoulos1997book} 
is employed to build feedback control laws mapping 
the outputs of the system (sensor signals) to its inputs (actuation commands).
MLC is  a function optimizer able to optimize both the structure of the control law and its parameters.
In an evolutionary process, new mechanisms (exploration) are found and 
are improved  (exploitation).
MLC has been successfully applied in dozens of experiments, each time outperforming optimized control methods 
often by exploiting unexpected nonlinear mechanisms \citep{Noack2019springer}.
MLC achievements include drag reduction of the Ahmed body with and without yaw angle \citep{Li2019prf,Li2018am}, 
jet mixing enhancement \citep{Zhou2020jfm} and mixing layer control \citep{Parezanovic2016jfm}, separation control of a turbulent boundary layer \citep{Debien2016ef}, recirculation zone reduction behind a backward facing step \citep{Gautier2015jfm}, reduction of vortex-induced vibration of a cylinder \citep{Ren2019pof,Ren2020jh} and pitch control for floating off-shore wind turbines \citep{Kane2020acc}.
Recently, MLC has been augmented with intermediate gradient descend steps for a fast descend into the minima \citep[gMLC]{CornejoMaceda2021jfm}.

This study constitutes, to the best of the authors' knowledge, 
the first self-learning model-free control for the stabilization of open cavity flows.
We employ our fastest optimizer, gMLC,  
to address the challenge of robust multi-frequency stabilization.
For this, feedback control laws are learned in two regimes: a narrow-bandwidth one and a mode-switching one.
The second regime constitutes a challenging problem as gMLC needs to learn a control law able
to control two modes simultaneously.
The robustness of the laws is tested by cross-evaluating each law in the other regime.

The manuscript is organized as follows.
\S~\ref{Sec:OpenCavityExperiment} introduces the cavity experiment setup including the wind tunnel, sensing, actuation and details the characteristics of the unforced dynamics.
 \S~\ref{Sec:ControlProblemAndMethodology} describes the control problem, 
 including the cost function and the ansatz for the control law, 
 and outlines  gradient-enriched machine learning control.
Moreover, two methods to interpret the control mechanisms are presented: 
an analytical approximation based on an affine regression and 
a cluster-based visualization method based on representative flow states.
In \S~\ref{Sec:Results}, the results of the control of the open cavity are described, 
from steady forcing as a benchmark to gMLC feedback.
\S~\ref{Sec:Discussion} discusses on the robustness of the gMLC laws, highlights the necessity for feedback and comments on the global nature of the achieved stabilization.
\S~\ref{Sec:Conclusions} summarizes the results and indicates directions for future research.
Table~\ref{tab:acronyms} lists all the acronyms used in the manuscript.

\begin{table}
\begin{center}
\def~{\hphantom{0}}
\begin{tabular}{p{1.75cm}p{8cm}}
\textbf{gMLC} & Gradient-enriched Machine Learning Control\\
\textbf{LDV} & Laser Doppler Velocimetry\\
\textbf{LGP} & Linear Genetic Programming\\
\textbf{MCS} & Monte Carlo sampling\\
\textbf{MDS} & Multidimensional Scaling\\
\textbf{MIMO} & Multiple-Input Multiple-Output\\
\textbf{MLC} & Machine Learning Control\\
\textbf{PSD} & Power Spectral Density\\
\end{tabular}
\caption{\label{tab:acronyms}Table of acronyms.}
\end{center}
\end{table}
 

%% file: S2.tex
\section{The open cavity experiment}
\label{Sec:OpenCavityExperiment}
This section details the characteristics of the wind tunnel, the means of sensing and actuation, the control unit and finally the unforced dynamics for the two regimes studied in this manuscript: the narrow-bandwidth regime and the mode-switching regime.

\subsection{Wind tunnel set-up}
\begin{figure}
\centering
\includegraphics[width=0.65\linewidth]{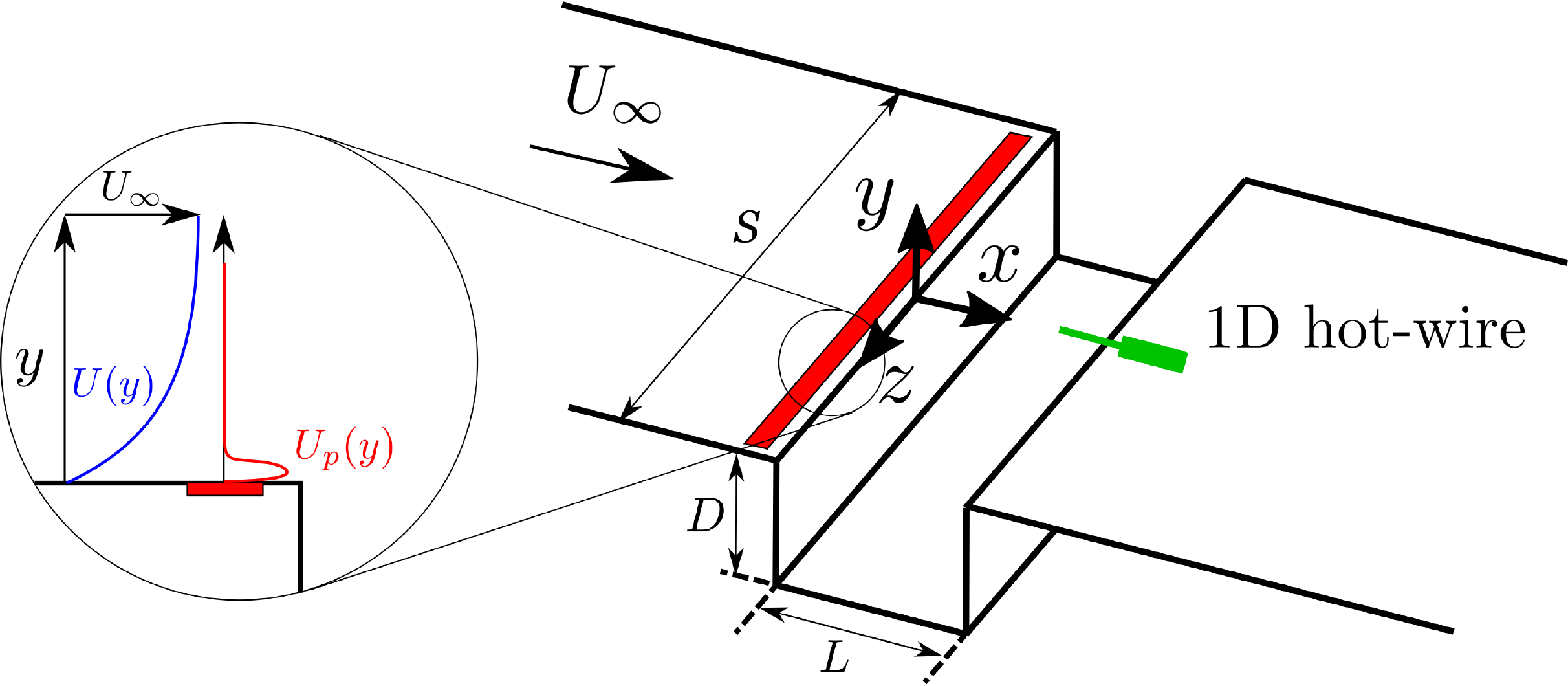}
\caption{Diagram of the cavity with the position of the DBD actuator (in red) and the velocity sensor (in green).
The magnified region depicts the velocity profiles of the incoming velocity ($U(y)$ in blue)
and the ionic wind produced by the DBD actuator ($U_p(y)$ in red).}
\label{fig:exp_setup}       
\end{figure}

The cavity is inserted into the rectangular cross-section duct of a $\SI{0.075}{\m}$ high and $\SI{0.30}{\m}$ wide  wind tunnel.
 The cavity, inserted in depression to the floor, is $D = \SI{0.05}{\m}$ deep, $S=\SI{0.30}{\m}$ wide and $L = \SI{0.075}{\m}$ or $L = \SI{0.0875}{\m}$ long following the studied regime.
 The resulting aspect ratio are $R = L/D = 1.5$ for the narrow-bandwidth regime and $R = L/D = 1.75$ for the mode-switching regime.
 A schematic of the wind tunnel is depicted in figure~\ref{fig:exp_setup};
 The walls are made of anti-reflection treated glass.
A Blasius-type boundary layer develops from an elliptical edge located $\SI{0.30}{\m}$ upstream.
Laser Doppler Velocimetry (LDV) measurements of the velocity upstream the cavity show that the standard deviation of the incoming flow is less than 1\%.
 
An anemometer is located at the exit of the open wind tunnel vein. 
Measurements show that the free-stream velocity $U_{\infty}$ and the velocity measured at the exit of the tunnel vein are linearly related to the rotation speed of the wind tunnel fan motor.
Thus in this study, $U_{\infty}$ is estimated from the anemometer measurements.
For the narrow-bandwidth regime, the incoming velocity is set to $U_{\infty} = \SI{2.13}{\meter\per\second}$, resulting on a Reynolds number equal to $\Rey_L = 1.04 \times 10^4$.
The momentum boundary layer thickness is estimated at $\theta_0/L=1.17 \times 10^{-2}$.
Great care has been taken to calibrate and regulate the incoming velocity with reference to LDV measurements.
However, it has been observed a 2\% variation of the incoming velocity for the narrow-bandwidth regime over the 24 hours necessary for the longest learning sessions.
The velocity variations are caused by the temperature variation, $T\approx\SI{23.14}{\pm2 \celsius}$, and very low cycle frequencies in the wind tunnel at this low velocity operating point.
The incoming velocity variations reach 5\% for the mode-switching regime.
Finally, the flow is in incompressible range with a Mach number less than $10^{-2}$.
A more detailed description of the set-up can be found in \citet{Lusseyran2008pof, Basley2013}.

\subsection{Hot-wire sensor}\label{Sec:HW_sensor}
For sensing, we use a constant temperature anemometer (DANTEC hot-wire probe 55P16 and miniCTA54T30 converter) with a single 1D hot-wire sensor, $\SI{5}{\micro\meter}$ in diameter and $\SI{1}{\milli\meter}$ length.
The hot-wire is located at $\SI{6}{\milli\meter}$ above the cavity and $\SI{6}{\milli\meter}$ upstream of the trailing edge, as sketched in green in figure~\ref{fig:exp_setup} and figure~\ref{fig:HotWire}.
The position of the hot-wire sensor has also be chosen to limit the velocity drops in the mode-switching regime, see \S~\ref{Sec:Unforced_dynamics}.
The hot-wire output signal from $E_{w}(t)$ is converted into streamwise velocity information $u$ according to King's law:
\begin{equation}
E_{w}^2 = A + Bu^n 
\end{equation}
where $A = 1.28$, $B = 0.70$ and $n = 0.48$  are determined by calibration of the hot-wire using an LDV anemometer.
Before conversion, the signal $E_{w}$ is temperature-corrected by the multiplicative factor $(T_w-T_0)/(T_w-T) $, where $T$ is the room temperature, $T_0$ is the calibration temperature and $T_w$ is the wire temperature \citep{Dantec2005}.
$T$ and $T_0$ are both measured with a $\rm P_{\rm t100}$ platinum sensor with $\SI{}{0.02 \celsius}$ accuracy.
The velocity measured $u$ is then employed in three ways: first, it serves to compute the performance of the tested controllers (\S~\ref{Sec:OptimizationProblem}); second, it closes the feedback control loop (\S~\ref{Sec:ControlProblem}); third, it is used to analyze the control mechanisms (\S~\ref{Sec:ControlLawInterpretation}).
All the following spectra and spectrograms are computed from this velocity measurement.

\subsection{Plasma actuator}
\begin{figure}
\centering
\begin{subfigure}{.45\textwidth}
  \centering
\includegraphics[width=0.8\linewidth]{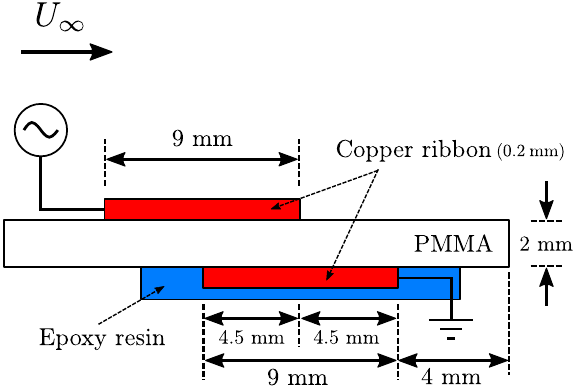}
\caption{Sketch of the DBD actuator.}
\label{fig:DBD_electrode}   
\end{subfigure}%
\hfil	
\begin{subfigure}{.45\textwidth}
  \centering
\includegraphics[width=0.8\linewidth]{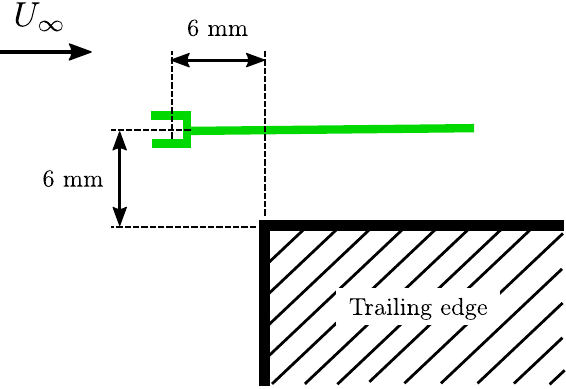}
\caption{Sketch of the hot-wire position.}
\label{fig:HotWire}   
\end{subfigure}
\caption{\label{fig:ActuatorSensor} The actuation is performed with a DBD actuator placed upstream (a) and the current state of the flow is given by a hot-wire probe downstream (b).}
\end{figure}

The actuation is carried out with a dielectric barrier discharge (DBD) actuator to locally force the boundary layer at the entrance of the cavity, near the separation edge where the receptivity of the shear layer is maximum \citep{Cattafesta1997aiaa} (see figure~\ref{fig:exp_setup}).
The DBD consists of two conductive blades placed on either side of an insulating plate and subjected to a high alternating voltage.
The streamwise shift between the two blades (see figure~\ref{fig:DBD_electrode}) creates an electric field parallel to the plate and responsible for an ionic wind in the streamwise direction.
The principle and the adjustment of the parameters for an application as a fluid actuator are thoroughly detailed in \citet{Moreau2007,Forte2007,Benard2010}.
In our experimental setup, the dielectric is made of $\SI{2}\mm$-thick acrylic glass (PMMA) and the electrodes are made of $\SI{9}\mm$-wide, $\SI{26}\cm$-long and $\SI{200}{\micro\meter}$-thick copper ribbons.
The downstream edge of the lower electrode is placed at $x=\SI{4}\mm$ upstream to the leading edge, see figure~\ref{fig:DBD_electrode}.

To produce an ionic wind, a carrying signal $E(t)$ at high frequency $f_p$ ($\approx \SI{3}{\kilo\hertz}$) is sent to the active electrode.
The signal $E(t)$ is produced by an agilent function generator and amplified ($\times 3000$) by a Trek high-voltage amplifier.
The expression of the carrying signal is:
\begin{equation}
E(t) =A(t) \> \sin(2\pi f_pt)
\end{equation}
with $A$ being the amplitude of the carrying signal.
The control is then achieved by modulation of the amplitude $A$ through the actuation command $b \in [-1,1]$.
In practice, $A$ is an affine function of $b$ such as $A|_{b=-1} = A_{\rm min}$ and $A|_{b=1}=A_{\rm max}$.
$A_{\rm min}$ is the ionization voltage; It is the threshold above which an ionic wind is produced.
The generated wind acts then as a localized body force whose intensity increases with the voltage and thus with $b$. 	
The increasing level of the body force results in the reduction of the main peak of the power spectrum until the dynamics are completely modified.
A steady actuation forcing study of the open cavity flow is reported in \S~\ref{Sec:SteadyActuation}.
$A_{\rm max}$ is defined as the maximum voltage that keeps the main resonance of the cavity still present in the power spectrum.

In practice, $A_{\rm min}$ and $A_{\rm max}$ are measured before each experiment as they are sensible to the atmospheric pressure, room temperature, moisture and number of hours of use of the electrode.
To make the control robust against these variations, the range of the actuation command $b$ is set independent of $A_{\rm min}$ and $A_{\rm max}$.

\citet{Forte2007, Moreau2007} describe the typical velocity profile generated by a DBD actuator with LDV measurements.
In particular, \citet{Forte2007} show that for a voltage of $\SI{20}{\kilo\volt}$ and a carrying frequency of $\SI{1}{\kilo\hertz}$ applied between $\SI{0.1}{\milli\metre}$-thick, $\SI{20}{\centi\metre}$-long aluminum electrodes, the velocity profile displays a maximum at $y=\SI{0.5}{\milli\meter}$ from the wall.
As the velocity profile moves downstream, the value of the maximum velocity decreases and its height increases up to $\sim \SI{1}{\milli\metre}$.
For our experiment, Pitot measurements indicate that for a tension equal to $\SI{6}{\kilo\volt}$ the maximum velocity is around $\SI{0.8}{\metre\per\second}$ and is reached at $y=\SI{1.25}{\milli\metre}$ of the wall.
Unfortunately, the tension value is not significative as $A_{\rm min}$ and $A_{\rm max}$ have changed by a factor between two experiments.

\subsection{Control unit}
\label{ControlUnit}
In our experiment, the signal acquisitions and actuation command are carried out by a dSPACE real-time controller, including a DS1600 4 cores processors board and a DS2201 I/O board with a 12 bits on $\SI{\pm10}{\volt}$ range analog-to-digital converter.
Only two inputs of the I/O board are exploited, one for the hot-wire signal and one for the voltage delivered by the $\rm P_{\rm t100}$ platinum sensor.
The hot-wire signal $E_w$ is translated and amplified ($\times40$) before analog-to-digital conversion.
All signals are sampled at $\SI{250}{\hertz}$ such as the Nyquist-Shannon theorem is respected up to three times the highest frequency of interest $f^+\approx \SI{40}{\hertz}$, also avoiding aliasing of the second harmonics.
One output of the I/O board is employed to send the command signal to the Agilent Function Generator.

The control optimization process includes two loops: a fast evaluation loop and a slow learning loop, see figure~\ref{fig:diag_mlc}.
The fast evaluation loop is managed by the ControlDesk software and Simulink.
For our study, the evaluation loop operates at the sampling frequency ($\SI{250}{\hertz}$).
For each control law tested, the time series of the actuation command and the hot-wire signal are recorded and post-processed with MATLAB.
As for the slow learning loop, it includes the post-processing of the control and the control law update;
It is automated with Python and MATLAB scripts.
Finally, the whole control unit is supervised by a PowerShell script that automates all the steps of the control optimization.

\subsection{Unforced dynamics}\label{Sec:Unforced_dynamics}
\begin{figure}
\centering
\begin{subfigure}{.5\textwidth}
  \centering
\includegraphics[width=\linewidth]{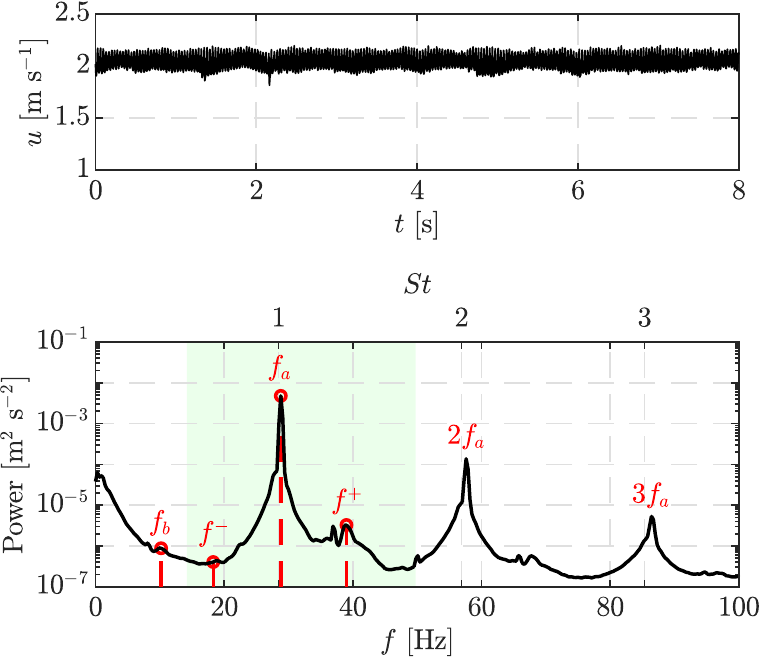}
\caption{Narrow-bandwidth regime}
\label{fig:NaturalMono}   
\end{subfigure}%
\hfil
\begin{subfigure}{.5\textwidth}
  \centering
\includegraphics[width=\linewidth]{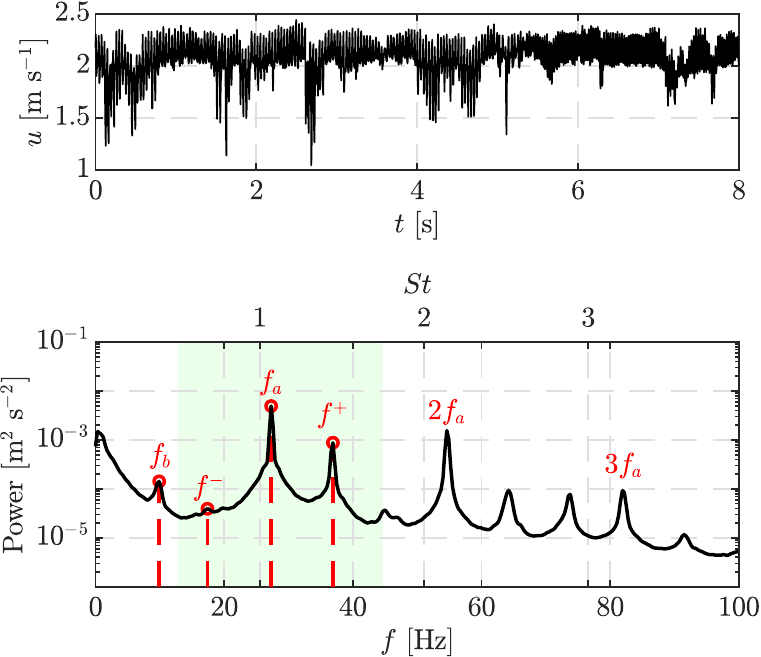}
\caption{Mode-switching regime}
\label{fig:NaturalIntermittent}   
\end{subfigure}
\caption{\label{fig:NaturalFlows} Time series (top) and spectral content (bottom) of the velocity measured downstream for the two unforced regimes.
The main frequencies of the flow are depicted in red: $f_b$, $f^-$, $f_a$, $f^+$ and $f_a$ harmonics.
The vertical axis of the spectra plots are in $\log_{10}$ scale.
The cost function (see \S~\ref{Sec:ControlProblem}) detects the maximum peak in the green-shaded window $\Str \in [0.5,1.75]$.
}
\label{fig:psd}
\end{figure}

As described in \S~\ref{sec:introduction}, the cavity allows a wide range of complex intra-cavity dynamics by tuning the two remaining cavity flow parameters namely the upstream speed $U_\infty$ and the width $L$.
We recall that the width $S$ and the depth $D$ of the cavity are fixed throughout this study and that the flow is incompressible (Mach number $< 10^{-2}$).
In this manuscript, we aim to stabilize two different flow regimes of different dynamical complexity.
For both regimes, the power spectrum is mainly organized within 5 frequency bands: the very low frequencies, not considered here, a low frequency $f_b$ and the 3 peaks directly reflecting the resonance of the mixing layer $f^-$, $f_a$ and $f^+$, see figure \ref{fig:NaturalFlows}.
These frequencies are nonlinearly coupled and satisfy the relationships $f^-= f_a - f_b$ and $f^+ = f_a+f_b$.
The two regimes studied differs by the power ratios of the frequencies $f_a$ and $f^+$.

The first regime is referred as the narrow-bandwidth regime and corresponds to a flow dynamics mainly centered on a single frequency $f_a$ and its harmonics.
This regime is achieved with $L=\SI{7.50}{\cm}$ and with an incoming velocity of $U_{\infty}=\SI{2.13}{\meter\per\second}$.
For this case, the ratio of the powers associated to $f^+$ and $f_a$ is close to $10^{-3}$, see figure~\ref{fig:NaturalMono}.
The coupling between $f_a$ and $f_b$ is then insignificant.
In contrast, for the second regime, referred as mode-switching regime, the power ratio between $f^+$ and $f_a$ is greater than $0.22$, see figure~\ref{fig:NaturalIntermittent}.
In this case, the nonlinear couplings between frequencies are strong and this leads to a chaotic intermittency between $f_a$ and $f^+$ \citep{Lusseyran2008pof}.
In the mode-switching regime, two modes compete in the flow leading to a switch of the dominant the frequency.
Such intermittency has been mentionned for the first time by \citet{Kegerise2004pof} for a compressible cavity flow.
In this study, the mode-switching regime is obtained in incompressible conditions ($\rm{Ma}<0.01$) for $L=\SI{8.75}{\cm}$ and for a slightly higher incoming velocity $U_{\infty}=\SI{2.23}{\meter\per\second}$, corresponding to a Reynolds number $\Rey_L=1.28\times10^{-4}$.
The momentum boundary layer thickness is estimated at $\theta_0/L=9.72\times10^{-3}$.
All the cavity flow parameters and experiment conditions are grouped in table~\ref{tab:paramRegimes}.

The episodic velocity drops, observed in the time series of the mode-switching regime (figure~\ref{fig:NaturalIntermittent}), are due to slow vertical undulations of the mixing layer which bring the low velocities of the lower part of the mixing layer to the level of the measurement point.
The position of the hot-wire sensor has been chosen to minimize these low-velocity incursions while limiting the damping of the oscillations to be controlled.
The undulations of the mixing layer are stronger for the mode-switching regime, and the incursions could not be avoided.

\begin{table}
\begin{center}
\begin{tabular}{>{\centering}p{2cm}>{\centering}p{1cm}>{\centering}p{3.5cm}>{\centering\arraybackslash}p{3.5cm}}
 Regime & Units & {\bf Narrow-bandwidth} & {\bf Mode-switching}  \\
\midrule
$U_\infty$ & $[\SI{}{\meter\per\second}]$ & 2.13 & 2.23\\
$L$ & $[\SI{}\cm]$ & 7.50 & 8.75 \\
$R=L/D$ &  & 1.5 & 1.75\\
$\Rey_L$ & & $1.04 \times 10^4$ & $1.28\times 10^4$ \\
$L/\theta_0$ & & 85.83 & 102.90\\
$T$ & $[\SI{}\celsius]$ & 22.60 &22.14 \\
$\nu$ & $[\SI{}{\square\meter\per\second}]$ & $1.53 \times 10^{-5}$ & $1.52 \times 10^{-5}$ \\
\end{tabular} 
\end{center}
\caption{Cavity flow parameters and experiment conditions for the two studied regimes.
The temperature $T$ and kinematic viscosity $\nu$ values are averaged over the learning session.}
\label{tab:paramRegimes}
\end{table}

\begin{figure}
\centering
\begin{subfigure}{.5\textwidth}
  \centering
\includegraphics[width=\linewidth]{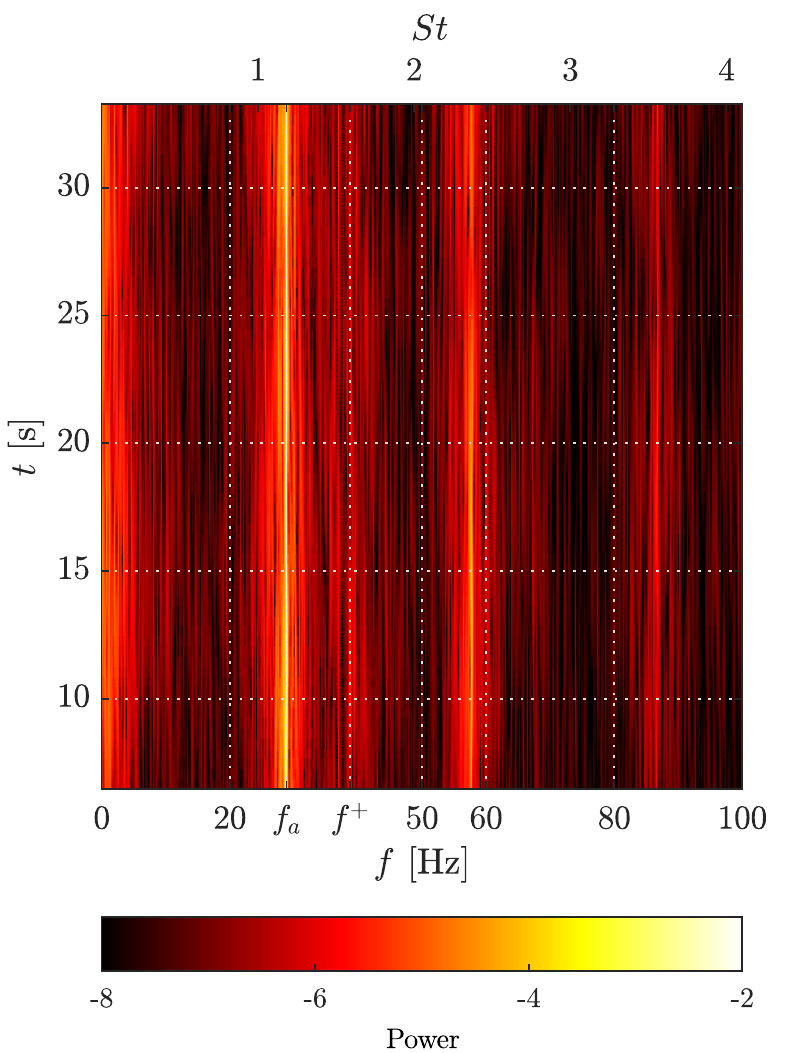}
\caption{Narrow-bandwidth regime}
\label{fig:SpectrogramMono}   
\end{subfigure}%
\hfil
\begin{subfigure}{.5\textwidth}
  \centering
\includegraphics[width=\linewidth]{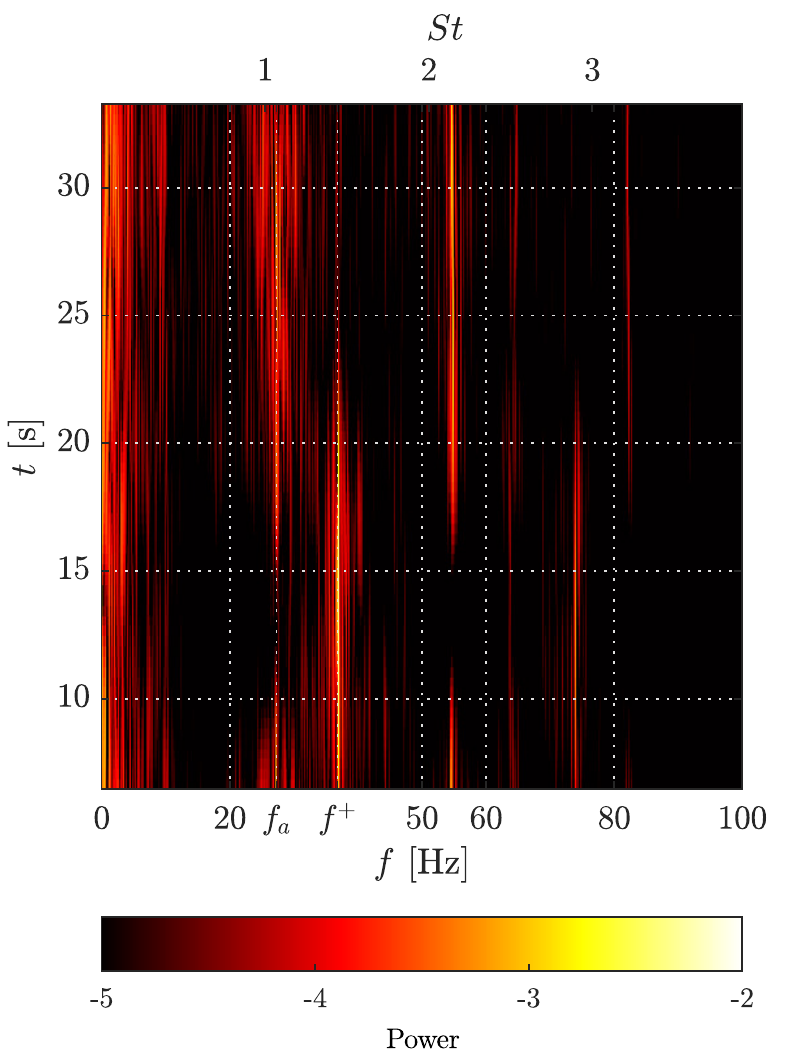}
\caption{Mode-switching regime}
\label{fig:SpectrogramIntermittent}   
\end{subfigure}
\caption{\label{fig:SpectrogramNatural}Spectrograms of the downstream velocity for the two studied regimes.
The color code corresponds to the power level in $\log_{10}$ scale.
The ticks of the color bar are the exposant values.
}
\end{figure}

The temporal evolution of the frequency content for the two regimes is depicted in figure~\ref{fig:SpectrogramNatural}.
In particular, figure~\ref{fig:SpectrogramMono} shows a clear line for the frequency $f_a$ and less intense lines for its harmonics, whereas figure~\ref{fig:SpectrogramIntermittent} displays a switching between frequencies $f_a$ and $f^+$ and their harmonics over the course of time.
The time between two switches is estimated between $\SI{15}{\second}$ and $\SI{20}{\second}$; Exceptionally, this time may exceed $\SI{40}{\second}$.

\begin{figure}
\centering
\includegraphics[]{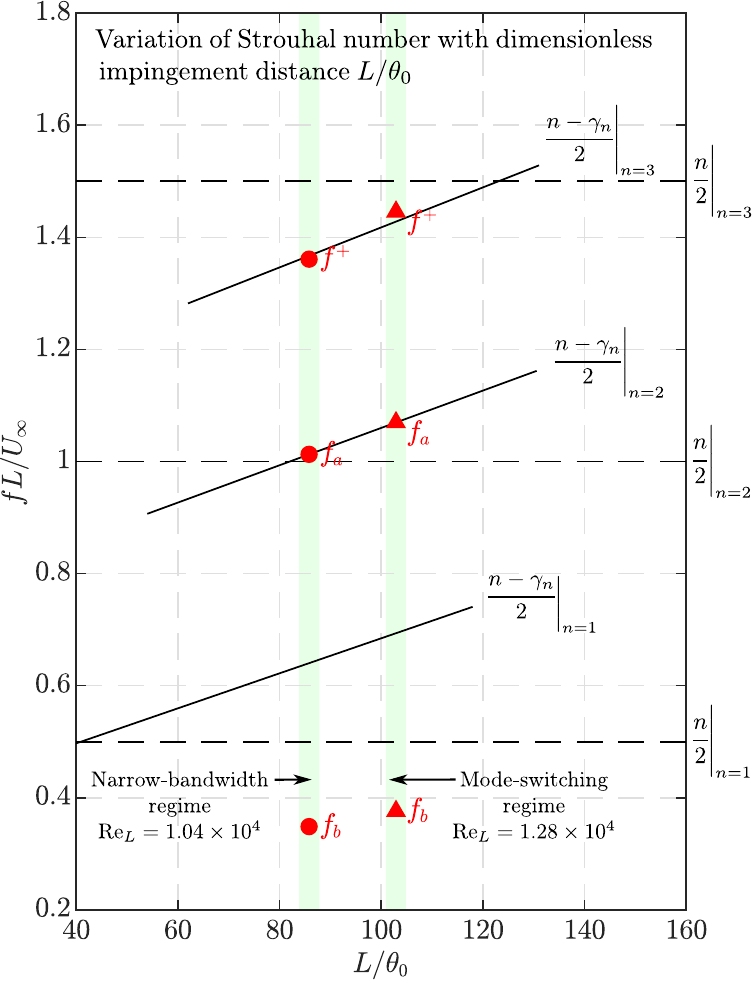}
\caption{Strouhal number ($St^{L}$) - dimensionless cavity length ($L/\theta_0$) map.
$\theta_0$ being the momentum boundary layer thickness and $L$ the cavity length.
Black lines represent the locus of equation~\eqref{eq:Str} for locked-on modes $n=1,2,3$.
Black dashed lines represent the locus of equation~\eqref{eq:Str} without the corrective term $\gamma_n$: $St^{L}_{n}=n/2$.
The left green band corresponds to the operating point for the narrow-bandwidth regime ($L/\theta_0=85.83$).
The right green band corresponds to the operating point for the mode-switching regime ($L/\theta_0=102.90$).
The red dots symbolize the three main oscillation modes of the mixing layer.
For more details see appendix~\ref{appB}). 
}
\label{fig:fromBasley2103}
\end{figure}

To understand the difference in dynamics between the two regimes, we locate them in the Strouhal versus $L/\theta_0$ map (figure~\ref{fig:fromBasley2103}).
Similar maps have been plotted for different impinging shear flows revealing jumps between the modes and linear-like relationships between the Strouhal number and the dimensionless cavity length $L/\theta_0$ or $L/\delta_0$, $\delta_0$ being the boundary layer thickness \citep{Sarohia1977,rockwell1978,Knisely1982}.
Indeed, \citet{Basley2013} shows that in such incompressible flow, most main frequencies measured in the downstream shear layer align with lines of locked-on modes such that the Strouhal number based on $L$ is given by
\begin{equation}
 St^{L}_{n} (L/\theta_0)=\frac{f_n\;L}{U_\infty}=\frac{n-\gamma_n(L/\theta_0)}{2},
\label{eq:Str}
\end{equation}
where the parameter $n = 1, 2, 3$ can be seen as the number of cycles within the cavity length and the corrective term, $\gamma_n$, can be interpreted as a wave adaptation to the effective resonance length.
The authors also propose a model for $\gamma_n$, linear with respect to the dimensionless cavity length $L/\theta_0$:
\begin{equation}
\gamma_n(L/\theta_0)=\frac{41n-L/\theta_0}{10\left(17-n\right)}.
\label{eq:gamma}
\end{equation}
On the other hand, it is noted that equation~\eqref{eq:Str} presents a resemblance with Rossiter's formula for compressible flows in \citet{Rossiter1964} where the corrective term is associated with the propagation time of the acoustic waves.

The Strouhal distribution is well described by \citet{Basley2013}, however it is worth noting that there is still no consensual overview for the origin of the incommensurable frequencies in incompressible open cavity flows.
As a first interpretation, the peaks in the spectrum are the result of nonlinear interactions inside the mixing layer dynamics.
Indeed, the resonance of the cavity occur for regimes beyond a critical Reynolds number $\Rey_c$ contrary to the Kelvin-Helmholtz instability of the mixing layer that is unstable  for all Reynolds numbers.
Beyond $\Rey_c$, self-sustained oscillations appear.
This scenario coming from a 2D perspective is however, a little simplistic when considering a real 3D cavity, especially in an incompressible regime at moderate Reynolds number.
In fact, the mixing layer develops in the streamwise direction and the flow not being strictly parallel,
Squire's theorem fails:
It is the transverse centrifugal instabilities that transit first, possibly several times, below $\Rey_c$. Depending on the values of the aspect ratio, these G\"{o}rtler-Taylor type instabilities have already reached, several bifurcations then a strong non-linear development when the resonant transition appears. This description corresponds to the two regimes chosen to test our control methodology and leads to the spectral signature described previously and in appendix~\ref{appA}.

In addition, \citet{Sipp2007jfm,Meliga2017jfm} have shown, using a linearization around the average flow in 2D simulations, that the occurrence of self-sustaining instabilities in shear-driven cavities are due to a supercritical Hopf bifurcation.
Following a similar method, \citep{Bengana2019jfm} and \citep{Tuerke2015prf} manage to predict two incommensurable frequencies in simulations and experiments respectively.
Another approach based on the identification of two characteristic delay times is able to predict the two frequencies of the flow \citep{Tuerke2020prf}.
The authors show that the non-linear interactions between these two frequencies can be captured by the resolution of a Stuart-Landau type amplitude equation, whose quadratic damping term consists of two delayed amplitude terms.
In this equation, the first delay time characterizes the upstream traveling hydrodynamic instability wave and hence the feedback of the reflected shear layer instability \citep{Tuerke2015prf}.
The second delay time is motivated by the hydrodynamic feedback of the recirculating vortices, also referred as ``vortex carousel'' and corresponds to an intra-cavity overturning time \citep{Tuerke2017a}.
In the following the description of \citet{Basley2013} that leads to figure~\ref{fig:fromBasley2103} is sufficient to guide the choice of parameters leading to the two regimes we have chosen to control.


Figure~\ref{fig:fromBasley2103}, plots the values of the measured frequencies in Strouhal number versus the dimensionless cavity length, for the two regime and their relation to the resonance points.
The details of computation of the momentum boundary layer thickness $\theta_0$ are detailed below.
For the first regime, $f_a$ is close to Strouhal number equals to $\left.n/2\right|_{n=2}$ i.e. at the intersection between the black line and the dashed line, while $f^+$ is clearly below the resonance at $n=3$.
As for the second regime, the Strouhal number corresponding to $f_a$ is above the resonant mode $n=2$ ($\gamma_2=-0.14$) and the one corresponding to $f^+$ is below the resonant mode $n=3$ ($\gamma_3=+0.14$).
In practice, $U_\infty$ has been chosen such as the average presence rate of the two frequencies $f_a$ and $f^+$ is equalized.
The fact that $\lvert \gamma_2 \rvert \approx \lvert \gamma_3 \rvert$ appears only after calculation, shows clearly that the parameter guiding the relative intensity of the two main modes is indeed $\lvert \gamma_n \rvert$.
The values of Strouhal number and $\gamma_n$ for each frequency are grouped in table~\ref{tab:freq}.

\begin{table}
\begin{center}
\def~{\hphantom{0}}
\begin{tabular}{>{\centering}p{2cm}>{\centering}p{4cm}>{\centering}p{1.5cm}>{\centering\arraybackslash}p{2.5cm}}
Peaks & $f_b\approx f^+-f_a\approx f_a-f^-$ & $f_a$ & $f^+$\\
\midrule
$n$ & - & 2 & 3 \\
\\
\multicolumn{4}{c}{\bf Narrow-bandwidth regime}\\

Freq. [\SI{}{\hertz}] & 9.91 & 28.81 & 38.72 \\
$St_n^L$ & 0.348 & 1.01 & 1.37 \\	
$\gamma_n$ & - & -0.0255 & 0.266 \\
\\
\multicolumn{4}{c}{\bf Mode-switching regime}\\
Freq.[\SI{}{\hertz}] & 9.60 & 27.31& 36.91\\	
$St_n^L$ & 0.376 & 1.07 & 1.43 \\
$\gamma_n$ & - & -0.140 & 0.143 \\
\end{tabular}
\end{center}
\caption{Characteristics of the regimes dynamics.
Values of the measured frequencies $f_b$, $f_a$ and  $f^+$ and their corresponding Strouhal numbers $St_n^L$.
The corrective term $\gamma_n$ is then computed from equation~\eqref{eq:Str}.}
\label{tab:freq}  
\end{table}

In fact, we observe a slight discrepancy between the natural frequencies measured and the predictions of \citet{Basley2013}, which we attribute to a change in the free development of the boundary layer and especially a reduction of the boundary layer thickness.
This reduction of the boundary layer thickness can be attributed to the planing effect of the $\SI{200}{\micro\meter}$ thick upper electrode, glued just before the leading edge.
Therefore, in this work, $L/\theta_0$ was not obtained from the Blasius law ($\theta_0=\kappa \sqrt{2 \nu l_x/U_\infty}$ with $\kappa=0.4696$, $l_x=\SI{0.3}{\meter}$) and $\nu$ the kinematic viscosity, nor by a direct measurement of $\theta_0$, for lack of optical access, but deduced from equations~\eqref{eq:Str} and \eqref{eq:gamma}, using the observed frequency (figure~\ref{fig:NaturalFlows}) and the regime parameters (table~\ref{tab:paramRegimes}) for the two considered regimes. 
First, the value of $\gamma_n$ is computed from $f_n$, $n$, $L$ and $U_{\infty}$ and equation~\eqref{eq:Str},
then $L/\theta_0$ is deduced from equation~\eqref{eq:gamma}.
The resulting dimensionless cavity lengths are $L/\theta_0=85.83$ for the narrow-bandwidth regime and $L/\theta_0=102.90$ for the mode-switching regime.

Finally, we have investigated the deviation obtained with the Blasius law.
From the values of $L/\theta_0$ and assuming the same expression as the Blasius law, we fit the corresponding kappa for our cases: for the narrow-bandwidth regime $\kappa=0.4209$ and for the mode-switching regime  $\kappa=0.4205$.
Both values are close to the value of the Blasius law ($\kappa=0.4696$) but slightly lower, which comforts the hypothesis of boundary layer thinning by the presence of the DBD electrode.

The low ($f_b$) and very low frequencies ($f<\SI{1}{\hertz}$) constitute a challenge for automatic learning as their more rare occurrences require longer time windows for converged statistics and thus slows down the overall learning process.
First, we have chosen to alleviate this difficulty by controlling the narrow-bandwidth regime where the very low frequencies ($f<\SI{1}{\hertz}$) are around two order of magnitude lower than $f_a$ in terms of power.
Then, we fully embrace the effect of the low frequencies with the mode-switching regime where the nonlinear interactions between $f_a$ and $f^+$ give rise to $f_b$ and especially the very low frequencies $f<\SI{1}{\hertz}$: $f_b$ is caused by the triadic interaction between $f_a$ and $f^+$ and the low frequencies ($f<\SI{1}{\hertz}$) are responsible for the frequency switches in the mode-switching regime.
Indeed, the power associated to the very low frequencies ($< \SI{1}{\hertz}$) is more than one order of magnitude greater for the mode-switching regime than for the narrow-bandwidth regime, see figures~\ref{fig:NaturalMono} and \ref{fig:NaturalIntermittent}.
The control of the low frequencies is then performed indirectly by controlling the two other frequencies $f_a$ and $f^+$.
Moreover, following \citet{Basley2014}, the energetic contribution of the very low frequencies is also due to the coupling between the mixing layer instability and the centrifugal instabilities originating in the span-wise direction within the cavity.

To conclude this description of the cavity dynamics, we recall that the goal we set for the control is to reduce the oscillation of the mixing layer by penalizing the peaks of power in the frequency range that includes $f^-$, $f_a$ and $f^+$ as indicated by the green shaded area of the figure~\ref{fig:psd}.


%% file: S3.tex
\section{Control problem formulation and methodology}
\label{Sec:ControlProblemAndMethodology}
In this section, the control problem is defined and the methodology to solve it and to analyze the solutions is described.
In \S~\ref{Sec:OptimizationProblem}, the control problem is reformulated as an optimization problem.
Such a problem is, in the most general case, non-convex and contains several minima a priori.
To solve such an intricate problem, we employ a powerful machine learning algorithm \S~\ref{Sec:gMLC}, the gradient-enriched machine learning control \citep[gMLC]{CornejoMaceda2021jfm}, that combines exploration to discover new minima and exploitation for a fast convergence.
Finally, two methods for describing the control mechanisms involved are presented: one based on linear regression and the second on the reconstruction of the phase space with clustering (\S~\ref{Sec:ControlLawInterpretation}).

\subsection{Cost function and optimization problem}\label{Sec:OptimizationProblem}
The aim of this study is the stabilization of the open cavity flow in two regimes of different complexity, in particular, the mitigation of the self-sustaining oscillations of the mixing layer.
For this, a cost function is built based on the velocity data provided by the hot-wire downstream.
The oscillations of the mixing layer are reflected in the oscillations of the velocity signal, thus the goal translates into the reduction of the highest peak of the associated power spectrum.
Moreover, the power invested and the power saved by the control must be balanced.
In that respect, two terms are considered in the cost function to optimize:
\begin{equation}
J = J_a+\gamma J_b
\label{eq:costfunction}
\end{equation}
The term $J_a$  accounts for the peak reduction and $J_b$ for the actuation power invested.
$J_{a}$ is defined as the value of the power spectral density maximum in a given frequency window.
The value is normalized by the value for the unforced case.
Hence, the performance of control law $K$ is given by:
\begin{equation}
J_a(K) = \frac{ \underset{\Str \in [0.5,1.75]}{\max} {\rm PSD}(u)}{\underset{\Str \in [0.5,1.75]}{\max} {\rm PSD}(u_0)}
\end{equation}
where ${\rm PSD}(u)$ is the power spectral density of the velocity $u$ measured by the hot-wire for the flow forced with the control law $K$ and $u_0$ is the velocity measured for the unforced flow.
A steady actuation forcing study (\S~\ref{Sec:SteadyActuation}) shows that the actuation affects both $f_a$ and $f^+$ so the detection window for the maximum of the PSD is set such as it comprises both $f_a$ and $f^+$: $\Str \in [0.5, 1.75]$.
Only the frequencies $f_a$ and $f^+$ are considered as they are the leading modes of the dynamics; The remaining high-power frequencies ($2f_a$, $3f_a$ in figure~\ref{fig:NaturalFlows}) are harmonics of the fundamental, i.e., slaved to $f_a$.
The detection window is set in Strouhal such as it is independent of the studied regime.
The normalization of the cost function $J_a$ by the value of the peak for the unforced flow allows us to have a direct measure of the reduction of the peak.

The $\rm PSD$ is computed over $T_{\rm ev}=\SI{40}{\second}$.
This choice is motivated for three reasons: First, it allows a good convergence of the statistics; Second, the time is short enough to evaluate 1000 individuals in few hours of experiment, limiting potential drifts and staying close to real-life applications with limited testing budget; Third, the mode-switching regime may include one or two switches during this period of time which is enough to have a record of both frequencies $f_a$ and $f^+$ in the spectrum.
Hence, the evaluation time balances practicality and good characterization of the flow dynamics.
Anticipating on the results, the value chosen for $T_{\rm ev}$ happened to be enough for the control of the two main frequencies in the mode-switching regime.
The control of the mode switching is realized indirectly by the control of the two frequencies involved $f_a$ and $f^+$.
It's worth noting that a direct control of the intermittency requires a much longer evaluation time due to its very low frequencies.

The actuation penalization term $J_b$ is estimated from the actuation command $b\in [-1,1]$, as the effective power supplied is not directly accessible in the experiment.
$J_b$ is based on the square of the actuation command averaged over $T_{\rm ev}$ so that it is an analogue to energy.
To simplify the interpretation, $J_b$ is normalized by the range of the actuation, so that $J_b=0$ when there is no actuation ($A=A_{\rm min}$) and $J_b=1$ when the controller acts steadily at maximum level ($A=A_{\rm max}$).
Therefore,
\begin{equation}
J_b = \frac{\langle (A-A_{\rm  min})^2 \rangle}{(A_{max}-A_{\rm min})^2} = \frac{\langle (b+1)^2 \rangle}{4}
\end{equation}
with $\langle . \rangle$ denoting the mean value over $T_{\rm ev}=\SI{40}{\second}$.
The choice of the penalization parameter $\gamma$ is based on the open-loop steady forcing study presented in \S~\ref{Sec:SteadyActuation}.
We show, in particular, that  a high level steady actuation is enough to reduce the cost  $J_a$ by at least $90\%$.
The penalization parameter $\gamma$ is chosen such as the cost for the unforced flow ($J_0=1$) is similar to the cost of the high level steady actuation (b =1), thus the optimal solution aimed needs to efficiently reduce $J_a$ with minimal actuation power.
As both cost function components $J_a$ and $J_b$ are normalized, we choose then the penalization parameter to be $\gamma=1$.
This choice results in setting the cost of the high level steady actuation (b =1) to $J(b=1)=J_a+J_b \approx 1.1\approx J_0$.

Finally, the normalized standard deviation $\widetilde{\sigma}$ of the velocity signal is computed for the best control laws, a posteriori, to characterize the controlled flow.
Indeed, an effective mitigation of the self-sustained oscillations of the mixing layer results in a reduction of the standard deviation, defined as such:
\begin{equation}
\widetilde{\sigma}(K) = \frac{\sigma (u)}{\sigma (u_0)}
\end{equation}
with $\sigma(u)$ being the standard deviation of the velocity $u$ computed over $T_{\rm ev}=\SI{40}{\second}$.
$T_{ev}$ is also chosen such as the standard deviation is sufficiently converged.
The standard deviation is normalized by the standard deviation of the natural unforced flow so to have a direct measure of the gain.

\subsection{Control problem}\label{Sec:ControlProblem}
As stated previously, the control objective is to stabilize the cavity flow by mitigating the oscillations of the mixing layer downstream.
To achieve this goal, the flow is forced with a DBD actuator located at the cavity leading edge.
The result of the action is an unsteady body force whose intensity is commanded by the input signal $b$ at the terminals of the DBD actuator.
$b$, also referred as the actuation command, is determined by the control law $K$.
The control may be open-loop or closed-loop with flow information input.
In this study, the considered open-loop actuations are only steady forcing and closed-loop control includes the unique velocity sensor and time-delayed records.
Thus, the control law reads:
\begin{equation}
b =K(\bm{a}) 
\end{equation}
with $\bm a$ being the feature vector comprising flow state information.
Then, the control problem to solve can be reformulated as an optimization problem where the goal is to derive the optimal control law $K^*$ that minimizes the cost function $J$.
\begin{equation}
K^* = \underset{K\in \mathcal{K}}{\operatorname{arg\,min}} \; J(K)
\label{eq:control_problem}
\end{equation}
with $\mathcal{K}:A \mapsto B$ being the space of all possible control laws.
$A$ is the input domain and $B$ is the output range for the actuation command.
Deriving the optimal control law $K^*$ without any a priori on the cost function $J$ is a challenging non-convex optimization problem presenting presumably several minima. 

\subsection{Gradient-enriched machine learning control} \label{Sec:gMLC}
In this section, we present the gradient-enriched machine learning control (gMLC) algorithm \citep{CornejoMaceda2021jfm} employed to solve the optimization problem~\eqref{eq:control_problem}.
Gradient-enriched MLC is an iterative function optimizer to derive control laws directly from the plant.
The method is based on machine learning control (MLC) \citep{Duriez2017book} and is augmented with downhill simplex steps to accelerate the learning.
MLC has already been employed to control dozens of experiments outperforming previous control laws with unexpected frequency crosstalk \citep{Noack2019springer}.
The choice of downhill simplex algorithm relies on its fast convergence and its easy implementation as it does not require an analytical expression of the cost function but only its evaluation.
In the past, downhill simplex has been successful in deriving an adaptive closed-loop control for lift-to-drag ratio optimization over a NACA 0025 airfoil \citep{Tian2006aiaa,Cattafesta2009chapter}, and reducing the net drag power of the fluidic pinball and a slanted Ahmed body \citep{LiA2022jfm}.
In \citep{CornejoMaceda2021jfm}, the gradient-enriched MLC is introduced and employed to successfully stabilize the fluidic pinball.
The authors show, in particular, that gMLC outperforms MLC, managing to derive better performing control laws with a greater learning speed.
It is now applied for the first time on an experiment.
Anticipating on the results (see \S~\ref{appA}), the superiority of gMLC over MLC is also verified for the control of the cavity in experimental conditions.
The benefits of gMLC compared to MLC comes from the combination of stochastic optimization for exploration of the search space and deterministic optimization for a fast convergence towards the minimum.
The methods consists on the generation of candidate solutions to equation~\eqref{eq:control_problem}, evaluate them and systematically recombine stochastically and deterministically the best ones to improve their performances.

Starting point of gMLC is MLC based on linear genetic programming \citep[LGP]{Brameier2006book}.
Following the genetic programming terminology, the candidate solutions are also referred as \textit{individuals}.
Like the MLC method, gMLC makes no assumptions on the structure of the relationship between the inputs and the outputs.
The optimal solution needs, however, to be computable, meaning it can be expressed by a finite number of mathematical operations with finite memory.
Indeed, the candidate solutions are internally represented by matrices inherited from linear genetic programming.
Each matrix resembles a computer program that unequivocally codes a control law.
Each line of the matrix is an instruction pointing to basic operations ($+$, $-$, $\times$, $\div$, $\cos$, $\sin$, $\tanh$, etc.) and registers containing constant random numbers and variables ($a_1$, $a_2$, $a_3$, etc).
The $N_{\rm inst}$ lines of the matrix are then read linearly yielding the control commands as outputs of the first registers.
We refer to \citep{Li2019prf} for more information on the internal representation of the control laws.

The gMLC algorithm starts with a broad exploration of the control law space with a Monte Carlo sampling (MCS) phase.
The Monte Carlo sampling generates $\Nmc$ random matrices that represent the first set of individuals.
The individuals are evaluated and added to the database of all individuals.
Then, the algorithm alternates between exploration phases carried out by genetic programming and exploitation phases performed by downhill simplex iterations until a stopping criterion is reached.
The role of exploration is to locate new and better minima in the space of control laws with a stochastic recombination of the best performing individuals.
The stochastic recombination is achieved with the genetic operations crossover and mutation.
This exploration is much like the evolution phase in the LGP method, however, in the case of gMLC, the concept of population that evolves through generations is generalized by considering all the individuals evaluated so far and stored in the database.
Thus, during the exploration phase, new individuals are generated by recombining the best among all the previously evaluated individuals.
This assures that no crucial information is irretrievably lost.
The best individuals to be recombined are selected following their cost function.
The selection is carried out by the tournament method with a tournament size equal to 7 for 100 individuals following \citet{Duriez2017book} recommendation.
The tournament size is scaled with the number of individuals in the database in order to keep a 7 for 100 ratio.
$\Nig$ new individuals are built at each exploration phase by recombining the best individuals of the database.

Each exploration phase is followed by an exploitation phase.
This step exploits the local gradient information to slide down towards the neighboring minimum.
This is carried out with a variant of downhill simplex for infinite-dimensional spaces introduced by \citet{rowan1990thesis} and referred to as downhill subplex.
In the following, we do not differentiate between the downhill simplex and subplex as the algorithms steps are similar and only applied to different spaces.
The principle of downhill simplex is to linearly combine the $N_{\rm sub}$ best-performing control laws following the gradient of the cost space to derive more performing individuals.
Contrary to the exploration phase, the new individuals are built in a deterministic way.
First, the $N_{\rm sub}$ best individuals are selected to describe a simplex that lives in the subspace generated by the $N_{\rm sub}$ best individuals.
The simplex, then, \emph{crawls} in the subspace according to geometric operations (reflection, expansion, contraction and shrink) following the local gradients.
Each geometric operation yields one or several new individuals that are linear combinations of the original $N_{\rm sub}$ individuals.
After each downhill simplex iteration, the simplex is updated by replacing the least performing individuals.
The downhill simplex steps are iterated until at least $\Nig$ individuals are generated.
The newly generated individuals are then added to the database of all individuals.
We emphasize that all the new individuals belong to the subspace defined by the original $N_{\rm sub}$ individuals.
If the stopping criterion is reached, the algorithm returns the best-performing control law. 
Otherwise, a new iteration of exploration and exploitation is carried out.
The stopping criterion may be a performance threshold or a total number of evaluations when the testing budget is limited.

\begin{figure}
\centering
\includegraphics[]{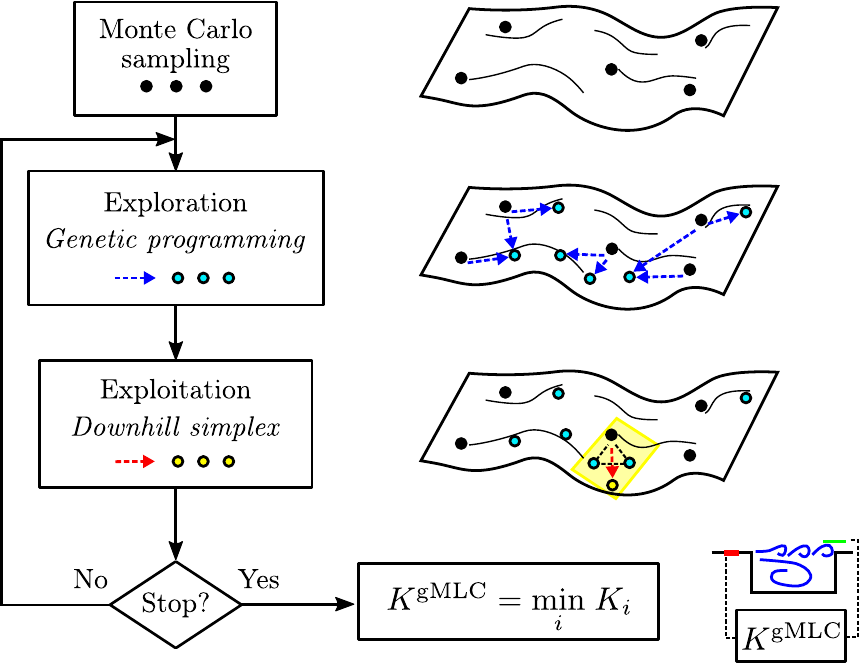}
\caption{Schematic of the gMLC algorithm.
On the left are depicted the three phases: Monte Carlo sampling, exploration and exploitation.
Each one of the phases requires several experiment evaluations.
On the right are depicted a representation of the control law space and the gMLC learning process.
The individuals generated during the Monte Carlo sampling are depicted in black.
The individuals resulting from genetic operations are depicted in blue and genetic operations by dashed blue arrows.
The yellow region represents the subspace built from the best individuals.
The individuals resulting from downhill simplex are depicted in yellow and the red arrows symbolize the simplex sub-steps (only one sub-step is depicted).
The reconstruction phase is not displayed for clarity.
}
\label{fig:gmlc_algo}
\end{figure}

We note the critical intermediate phase of reconstruction between each exploitation and exploration iterations.
Indeed, the new individuals generated by the downhill simplex are linear combinations of individuals without a matrix representation, which is essential for the genetic recombination during the exploration phase.
Thus, a matrix reconstruction is performed for each linearly combined individual by solving a secondary optimization problem.
The goal is to derive a matrix which translates into a control law that has the same response as the linearly combined one.
Such a problem is similar to a surface fitting problem, which we solve with linear genetic programming.
The reconstruction phase builds a matrix representation for the linearly combined individuals in such a way that they can be recombined with genetic operators.
For more information on the gMLC algorithm, we refer the readers to \citet{CornejoMaceda2021jfm}.
Figure~\ref{fig:gmlc_algo}, schematically illustrates the different phases of the gMLC algorithm and the learning principle in the control law space.
The MATLAB implementation of gMLC employed for this study is freely available at \url{https://github.com/gycm134/gMLC}.

\subsection{Control law investigation}\label{Sec:ControlLawInterpretation}
In this section, we propose two methodologies to analyze the actuation  mechanisms of optimized control laws.
Firstly, an analytical approximation of the control is performed with an affine mapping between the inputs (components of $\bm a$) and the actuation command ($b$).
Such mapping aims to reveal the most relevant component of the feature vector.
The affine approximation $\widetilde{K}$ of the control law $K$ reads:
\begin{equation}
\widetilde{K}(\bm a) = k_0 + \sum_{i}k_i a_i
\end{equation}
where $k_i$ are gains determined by linear regression between the time series $\bm a$ and $b$ recorded during the experiment.
The quality of the fitting is measured by the coefficient of determination $R^2$, measuring the relative reduction of the residual variance.
The closer $R^2$ is to 1, the better $\widetilde{K}$ fits the original control law $K$.

Secondly, we propose a visualization of the control laws based on the clustering of the feature vector $\bm a$ to reconstruct the phase portrait.
Cluster-based methods have been successful in reproducing key characteristics of fluid flow dynamics such as temporal evolution and fluctuation levels \citep{Fernex2021,LiH2021}.
For this analysis, all the states of the feature vector are grouped in 10 clusters to reconstruct the dynamics.
The cluster centroids, $c_k$, are defined as the average state of all the states in a given cluster.
Clustering is performed with the k-means algorithm \citep{Lloyd1982ieee} and the metric employed is the one induced by the $L^2$ norm.
The dynamics of the feature vector are then encapsulated in a probability transition matrix where its elements $p_{ij}$ are the transition probabilities from cluster $i$ to cluster $j$.
The probability $p_{ij}$ is defined as $p_{ij}=n_{ij}/n_i$ with $n_{ij}$ being the number of states transitioning from cluster $i$ to cluster $j$ and $n_i$ the total number of states in cluster $i$.

Then all feature vector states and centroids are projected on a two-dimensional space with classical multidimensional scaling \citep[MDS]{Kaiser2017ifac,LiA2022jfm}.
MDS is dimensional reduction method that consists on extracting the two main features of the flow ($\gamma_1$ and $\gamma_2$) by applying a proper orthogonal decomposition on the distance matrix of the feature vector $\bm a$.
The vectors $\gamma_1$ and $\gamma_2$ spawns a two-dimensional space where all the data is projected.
It is the optimal projection, in the $L^2$ norm sense, that preserves the distances between the states.
Such representation is referred as a proximity map.

\begin{figure}
\centering
\includegraphics[width=0.75\textwidth]{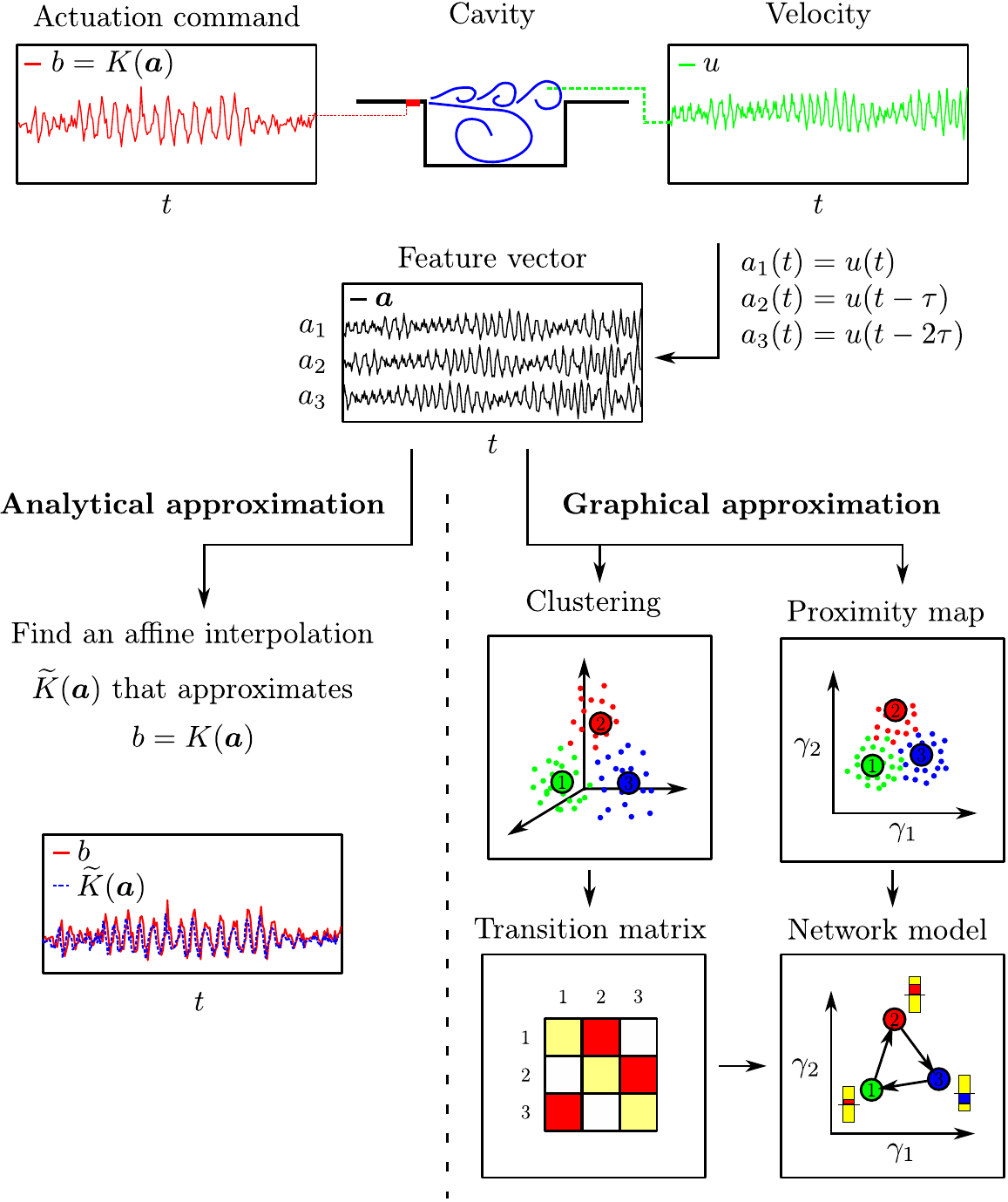}
\caption{Control law investigation methodology for the cavity control.
A feature vector $\bm a$ is built from a direct measurement of the flow $u$.
A feature vector of dimension three is depicted for simplicity.
The elements of the feature vector are grouped in clusters.
For clarity, only three clusters and their centroids (1,2,3) are displayed.
The proximity map is defined by $\gamma_1$ and $\gamma_2$ the two flow features extracted with classical multidimensional scaling.
For the transition matrix, darker squares symbolize higher transition probabilities.
In the control network model, the actuation magnitude are represented by rectangles; yellow denote the actuation range, red (blue) for a positive (negative) actuation with respect to the mean value.}
\label{fig:InterpMethodo}
\end{figure}

Adding the probability transitions to the proximity map allows to build a network model reproducing the phase portrait.
The centroids constitute representative states of the flow where the system transitions ergodically, meaning that from any centroid one can reach any other centroid.
Finally, the mean forcing level is computed for each cluster and associated to their corresponding centroid.
Such representation allows to visually partition the states of strong or low forcing and to reveal actuation mechanisms.
Figure~\ref{fig:InterpMethodo} summarizes the two approximation methodologies employed.

The latter data-driven methodology for control visualization is expected to aid the human interpretability of machine-learned controls.
In this study, the methodology is employed to analyze a single-input single-output system, however we believe that the methodology will be beneficial for the analysis of more complex control systems including a high number of inputs and outputs.

%% file: S4.tex
\section{Control results}\label{Sec:Results}
In this section, we stabilize the open cavity flow in two regimes: the narrow-bandwidth regime (\S~\ref{sec:gmlc_R1}) and the mode-switching regime (\S~\ref{sec:gmlc_R2}) presented in \S~\ref{Sec:Unforced_dynamics}.
We recall that the control objective is to mitigate the self-sustaining oscillations of the mixing layer.
First, in (\S~\ref{Sec:SteadyActuation}) we reduce the main oscillations with steady forcing at increasingly actuation level.
Then, we employ gradient-enriched machine learning control to optimize feedback control laws.
\S~\ref{sec:gmlc_settings} details the parameters employed for the control law optimization and \S~\ref{sec:gmlc_R1} and \S~\ref{sec:gmlc_R2} present the results for the control of the narrow-bandwidth regime and the mode-switching regime respectively.

\subsection{Open-loop steady forcing}\label{Sec:SteadyActuation}
\begin{figure}
\centering
\begin{subfigure}{.45\textwidth}
  \centering
\includegraphics[width=\linewidth]{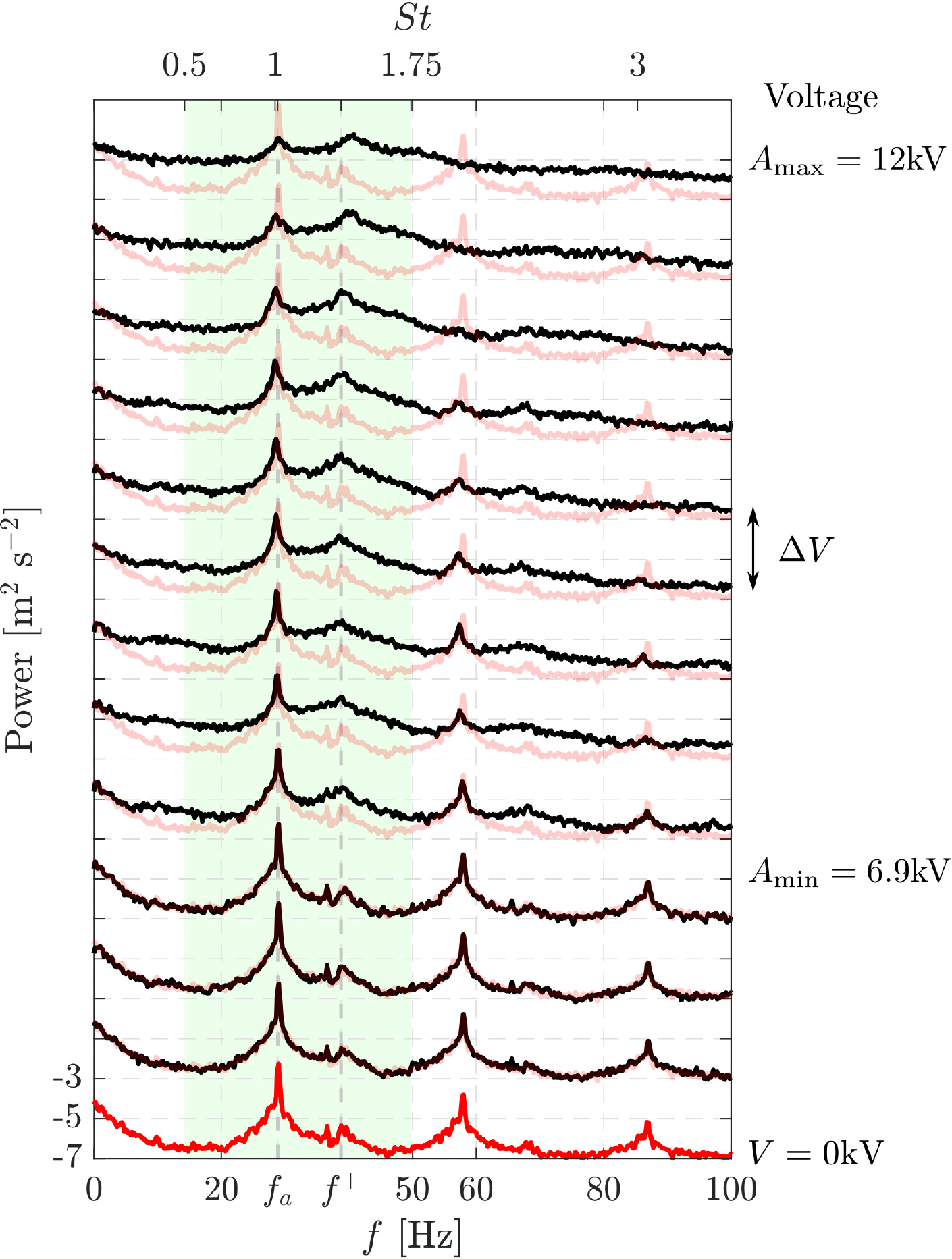}
\caption{Narrow-bandwidth regime}
\label{fig:SAResponseR1}   
\end{subfigure}%
\hfil
\begin{subfigure}{.45\textwidth}
  \centering
\includegraphics[width=\linewidth]{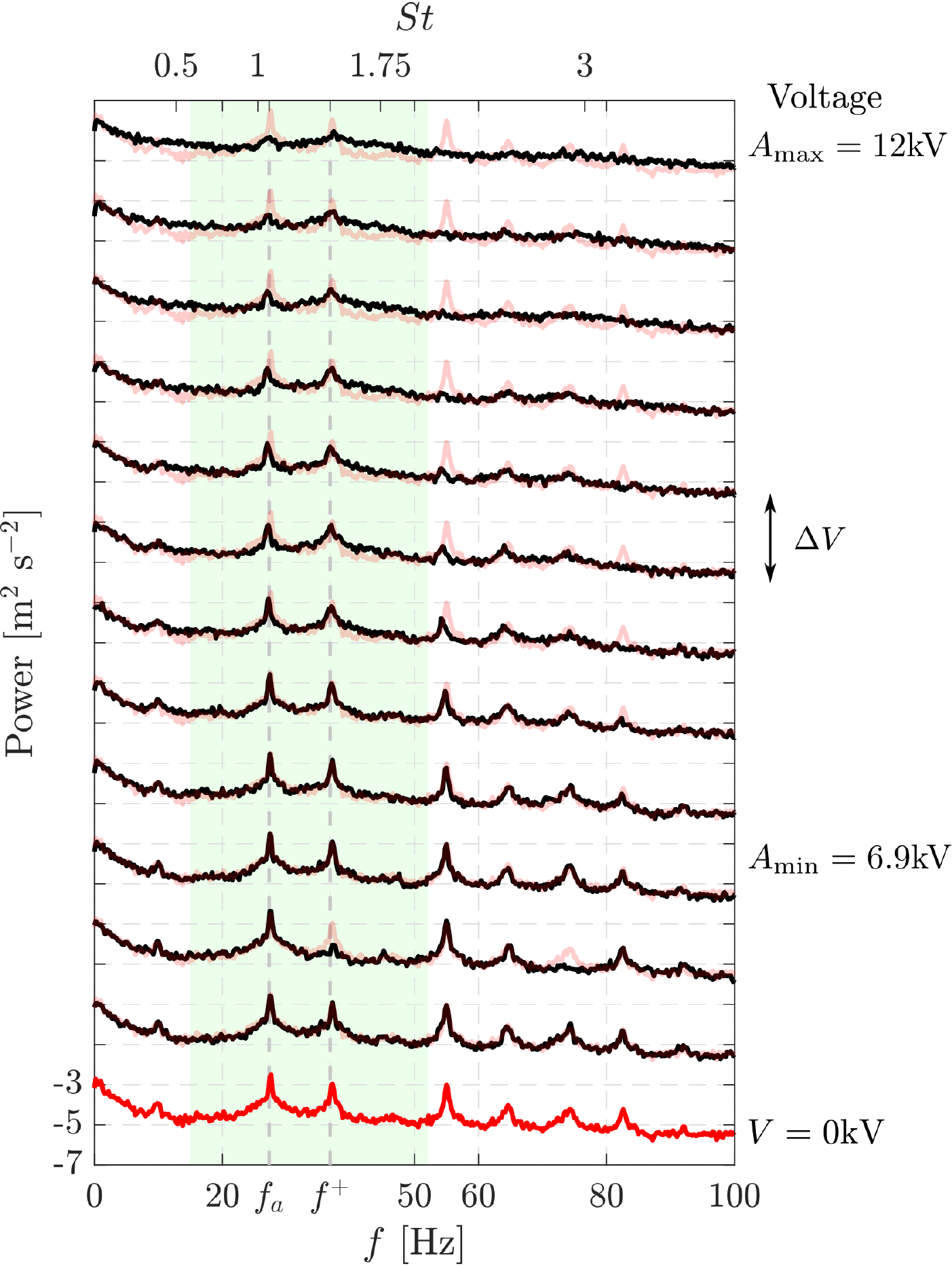}
\caption{Mode-switching regime}
\label{fig:SAResponseR2}   
\end{subfigure}
\caption{\label{fig:SteadyActuationResponse}Response of the cavity flow to increasing levels of steady actuations.
Only 12 actuation levels among the 23 tested are depicted for clarity.
The vertical axis are in $\log_{10}$ scale and the ticks are the exposant values.
The spectra have been successively shifted upwards of 4 orders of magnitude to help the comparison.
The difference of actuation intensity between two spectra is $\Delta V=\SI{0.6}{\kilo\volt}$.
The spectrum of the unforced flow is plotted in red and in transparency for comparison with the other levels.
}
\end{figure}

In this section, the response of the flow to steady actuations is described.
For this study, the amplitude of the carrying signal is set to constant values.
The flow is excited with 23  levels of actuation equally distant from the ionization level ($A=A_{\rm min}=\SI{6.9}{\kilo\volt}$) to $A=A_{\rm max}=\SI{12}{\kilo\volt}$.
Figure~\ref{fig:SteadyActuationResponse} presents the velocity power spectra for the two regimes.
For the narrow-bandwidth regime, figure~\ref{fig:SAResponseR1} shows that the second peak $f^{+}$ rises and the first peak $f_a$ decreases as the actuation level increases.
When the actuation is too strong, the noise level increases and the two peaks are at the same level.
The maximum peak reduction is achieved for $A=A_{\rm max}$ with a cost reduction of $97 \%$.
The associated standard deviation slightly decreases to $\widetilde{\sigma}=96 \%$.

For the mode-switching regime, figure~\ref{fig:SAResponseR2} shows that a quite strong actuation level is needed to reduce the amplitude of the peaks associated with $f_a$ and $f^+$, though the broadband noise level also increases.
The maximum cost reduction is achieved for the nearly maximum actuation level $A=\SI{11.4}{\kilo\volt}$ ($88\%$ of $A_{\rm max}$) and the maximum peak power decreases by $90\%$.
Also the standard deviation slightly decreases to $\widetilde{\sigma}=95 \%$.
We note that for the third spectrum starting from the bottom ($V=\SI{1.2}{\kilo\volt}$), the incidental absence of mode switching during the measurement led to a lower $f^+$ peak.

This open-loop analysis shows that a strong steady actuation is able to reduce the fluctuations of the shear layer.
However, in both cases the background noise level increases.
Therefore, in the following, in order to exclude power demanding controllers, we consider the cost function described in \S~\ref{Sec:OptimizationProblem} that includes two terms: one based on the maximum amplitude of the spectrum and an actuation penalization term.

\subsection{Implementation of gradient-enriched machine learning control}
\label{sec:gmlc_settings}
The parameters chosen for the gMLC algorithm are similar to the ones chosen in \citet{CornejoMaceda2021jfm}.
The Monte Carlo sampling phase generates $\Nmc=100$ individuals.
The exploration and exploitation phases both produce $\Nig=50$ new individuals at each iteration.
The exploration and exploitation phases alternate until 1000 individuals are evaluated.
We recall that each individual is evaluated over $T_{\rm ev}=\SI{40}{\second}$.
A relaxation time of $\SI{2}{\second}$ is intercalated between two consecutive control law evaluation.
The experiment time needs also to include the time needed to solve the reconstruction problem, but this constraint can be lifted with additional computation power.
The limit of 1000 individuals is then chosen such as all the individuals are evaluated in one day.
\citet{CornejoMaceda2021jfm} shows also that 1000 evaluations is enough to converge for a multiple-input multiple-output problem.
The subplex space is generated by $\Nsub=10$ control laws to balance speed and performance, as in \citet{CornejoMaceda2021jfm}.
For the evolution during the exploration phase, the crossover and mutation probabilities are both set to $\Pc=\Pm=0.5$.
The control laws are built from nine mathematical operations ($+$, $-$, $\times$, $\div$, $\sin$, $\cos$, $\tanh$, $\exp$ and $\log$), ten flow features $\{a_i\}_{i..10}$ and $\Ncst=10$ random constants.
As suggested by \citet{Duriez2017book}, the $\div$ and $\log$ operations are protected allowing them to be defined for all the real numbers.
$\Nvar=14$ registers are employed to derive the control laws.
Finally the maximum number of instructions to be coded in the matrix representation is $\Ninstrmax=50$.
The flow features employed for feedback control laws are the velocity signal and nine time-delayed velocity signals.
Time-delayed sensor signals are introduced as inputs to enrich the search space and allow, in principle, ARMAX type controllers \citep{Herve2012jfm}, linear and nonlinear combinations of them.
The resulting feature vector $\bm a$ reads:
\begin{equation}
\bm{a}(t) = [u(t),u(t-\tau),\ldots,u(t-9\tau)]^{\intercal}
\end{equation}
where the time delay is $\tau = \SI{0.008}{\second}$.
The choice of the delays allow to reconstruct the phase of the main frequencies of the flow between $\SI{25}{\hertz}$ and $\SI{40}{\hertz}$ \citep{CornejoMaceda2021jfm}.
Actually, only half of the delays are necessary but nine have been taken into account to enrich the phase space and get closer to full-state control.
The presence of time-delayed information in the control can play, for example, the role of an embedding process in the new dynamical system consisting of the flow and the closed-loop control.
Indeed, we have opted for single measurement point separated from the actuator by a convective time that intrinsically varies over time.
Implicitly, the system under loop control is hence reduced to a purely temporal dynamical system and the spatial information can be interpreted as an embedding of this dynamic into a larger phase space.

\begin{figure}
\centering
  \includegraphics[width=0.8\textwidth]{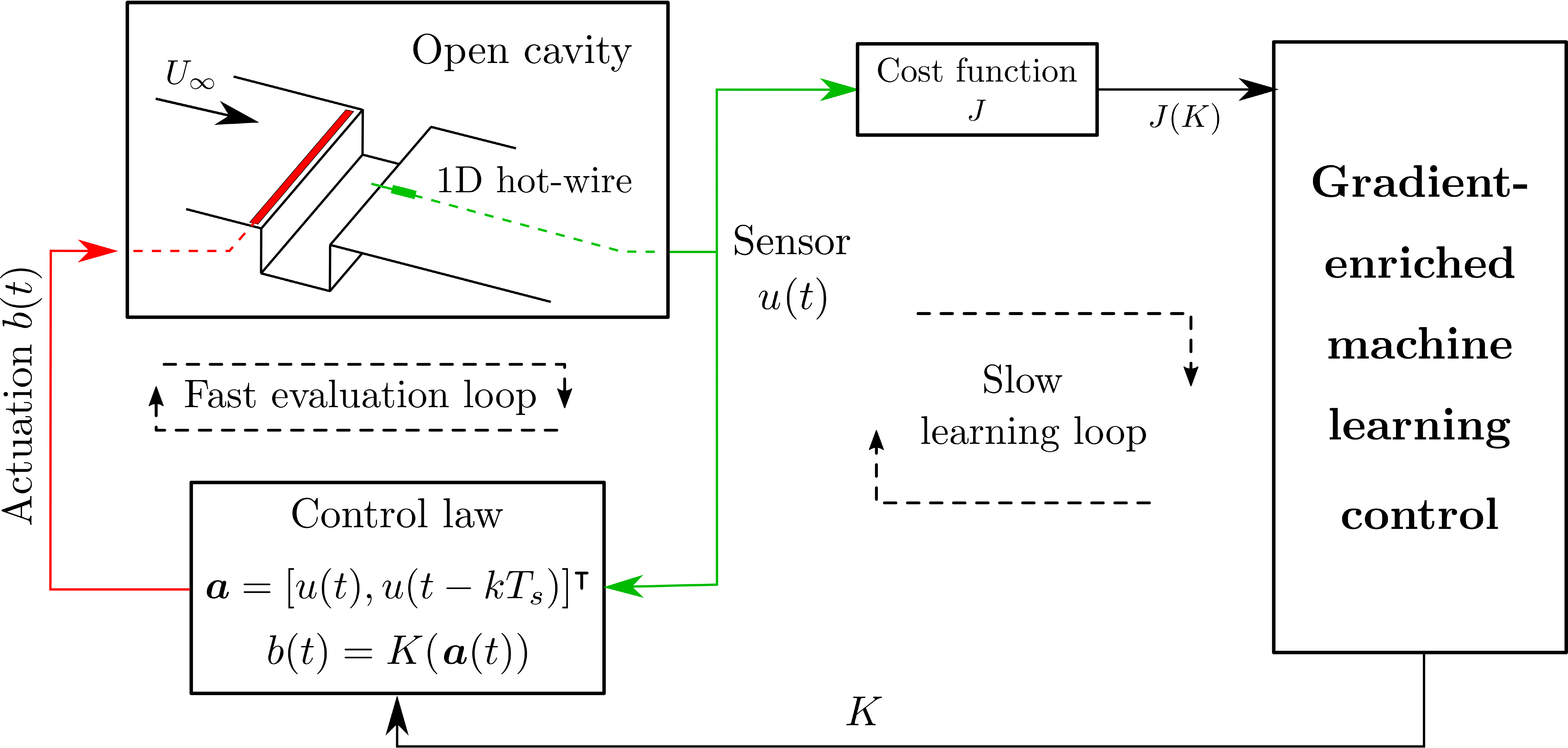}
\caption{Diagram of the machine learning control learning process.
The evaluation of one individual constitutes the fast evaluation loop, operating at $\SI{250}{\hertz}$.
The slow learning loop updates the control laws following the gMLC algorithm; it operates at $\sim O(10^{-2}\SI{}{\hertz})$.
}
\label{fig:diag_mlc} 
\end{figure}

Table~\ref{tab:cavity_parameters} summarizes the parameters for the gMLC optimization process.
\begin{table}
\begin{center}
\def~{\hphantom{0}}
\begin{tabular}{>{\centering}p{2.0cm}>{\centering}p{6cm}>{\centering\arraybackslash}p{5cm}}
  Parameter & Description & Value\\
\midrule
& Function library & $+$, $-$, $\times$, $\div$, $\sin$, $\cos$, $\tanh$, $\exp$, $\log$ \\
$\Nb$ & Number of controllers & 1\\
$\bm a$ & Control law inputs & $u(t),\ldots,u(t-9\tau)$\\
$\Nvar$ & Number of variable registers & 14 \\
$\Ncst$ & Number of constant registers & 10 \\
$\Ninstrmax$ & Max number of instructions & 50\\
$\Nmc$ & Number of Monte Carlo individuals & 100 \\
$\Nsub$ & Subspace size & 10 \\
$\Pc$ & Crossover probability & 0.5 \\
$\Pm$ & Mutation probability & 0.5 \\
$\Nig$ & Number of individuals per phase & 50 \\
\end{tabular}
\end{center}
\caption{\label{tab:cavity_parameters}Gradient-enriched MLC parameters to control the open cavity.}
\end{table}
Figure~\ref{fig:diag_mlc} displays the experiment setup to control the open-cavity with machine learning control and specifically with gMLC.

\subsection{Closed-loop control of the narrow-bandwidth regime}\label{sec:gmlc_R1}
In this section, we describe the best control law derived by gMLC that mitigates the self-sustaining oscillations of the mixing layer for the narrow-bandwidth regime.
In the following, the notations for the control law, cost and standard deviation associated with the learning on the narrow-bandwidth regime are marked by the superscript $\textrm{I}$.

\begin{figure}
\centerline{\includegraphics[width=0.90\linewidth]{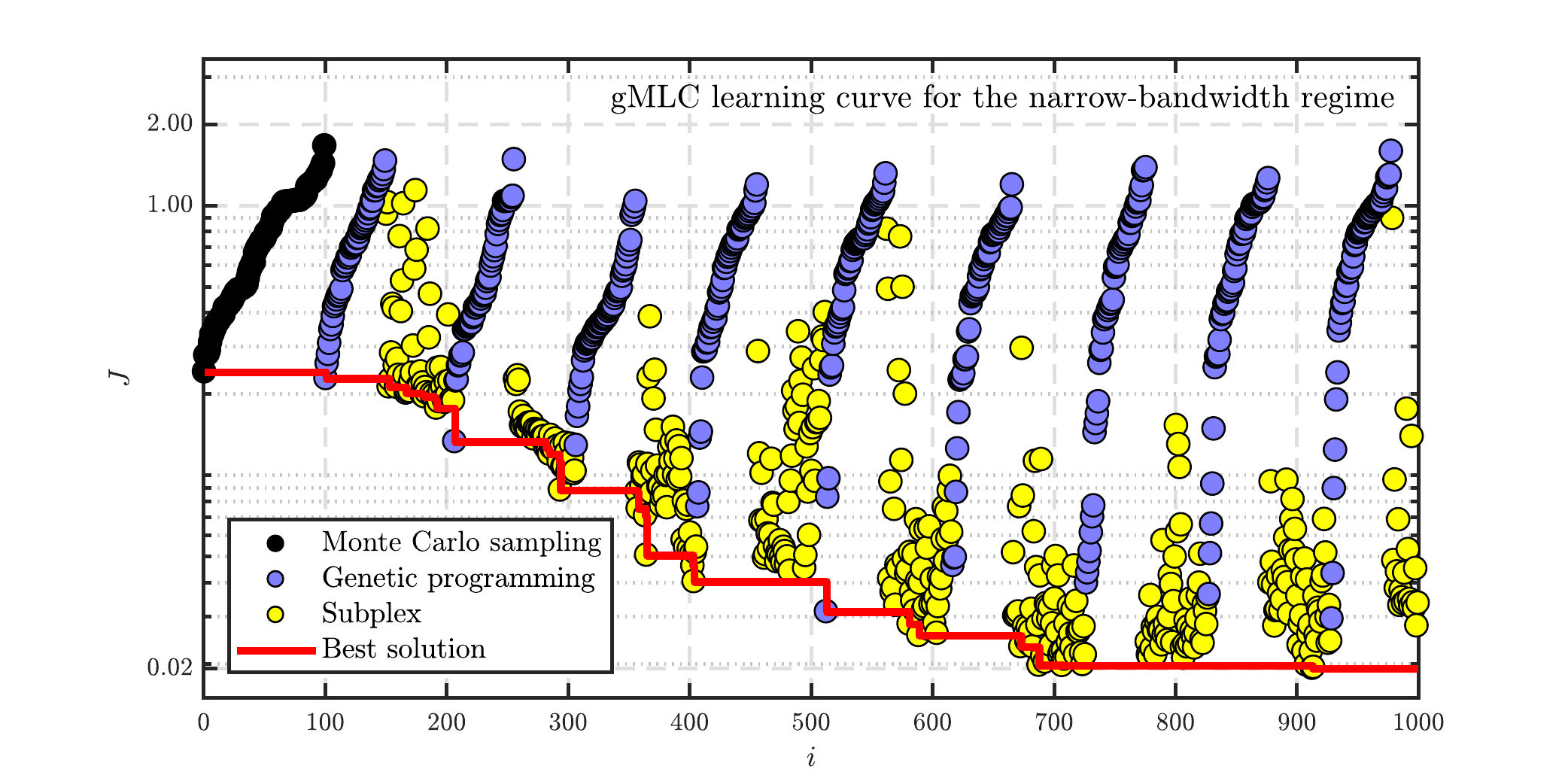}}
 \caption{Distribution of the costs for the 1000 evaluated individuals during the gMLC optimization process for the narrow-bandwidth regime.
Each dot represents the cost $J$ of one individual.
The color of the dots symbolize how the individuals are generated: 
black dots for randomly generated individuals (Monte Carlo sampling phase), blue dots for individuals generated by a genetic operator (exploration phase) and yellow dots for the individuals arising from the subplex method (exploitation phase).
The individuals from the Monte Carlo sampling and exploration phases are sorted following their costs.
The red line follows the evolution of the best cost.
The vertical axis is in log scale.}
\label{fig:ProgressR1}
\end{figure}

Figure~\ref{fig:ProgressR1} depicts the cost for the 1000 evaluated individuals during the optimization process.
The individuals from the Monte Carlo sampling and exploration phases are sorted following their cost as there is no direct causal relationship between two successive individuals,
while the individuals generated during the exploitation phase are depicted in the order of their evaluation as each individual depends of the previous one.
We recall that the cost function is defined such as the cost of the unforced flow is $J_0=1$.
We note that a random sampling of 100 control laws already manages to reduce the cost function to $J=0.2410$.
Then, the first exploration phase (individuals $i=101,\ldots,150$) reduces slightly the cost function to $J=0.2276$.
In the following exploitation phase (individuals $i=151,\ldots,206$), the downhill subplex individuals reduces gradually the cost function to $J=0.1882$.
We note that, at first, the individuals are scattered along the vertical axis and then go down, close to lowest cost so far.
This behavior shows that the individuals progress towards a minimum in the control law space.
However, this descent is interrupted by the next exploration phase (individuals $i=207,\ldots,257$) where a more performing individual is found, its cost is $J=0.1329$.
This new individual replaces the least performing individual in the simplex, allowing to explore beyond the initial subspace.
It is worth noting that the dimension of the subspace remains the same as new individuals replace the least performing ones.
From there on and until individual $i=512$, only the exploitation phases built better individuals.
The cost of the best individual after 512 individuals is $J=0.0402$.
Interestingly, we note that downhill simplex can build worse performing individuals.
Indeed, we notice that all exploitation phase starting from the $4$-th one include individuals whose costs are spread up to $J=1$.
The next exploration phase (individuals $i=513,\ldots,563$) finds a better control, whose cost is $J=0.0311$.
As the simplex includes now poor performing individuals, four better performing individuals generated by the exploration phase are introduced in the simplex.
From there on, progress is made only with exploitation steps: the cost of the best individual reaches a plateau after $707$ evaluations and is only slightly improved after 913 evaluations.
After 1000 evaluations, the cost of the final control law $\KgMLCRa$ is $\JgMLCRa = 0.0192$.
The corresponding peak reduction is $\JagMLCRa=0.0129$, i.e. 99\% of the fluctuation level
In other terms, the amplitude of the oscillations is reduced by a factor 9 or by $\SI{19}{\decibel}$.
The structure and components of $\KgMLCRa$ is thoroughly described in appendix~\ref{appC}.

The learned control law $\KgMLCRa$ is re-evaluated 20 times afterwards to test its efficiency outside the learning loop.
We note that the performances slightly dropped as the averaged cost reduction of $\JgMLCRa$ went from $-98\%$ to $-94\%$ and the standard deviation increased from $\siggMLCRa=61\%$ to $\widetilde{\sigma}=65\%$.
Such discrepancy is expected as the experimental conditions are always evolving.
Indeed, the temperature in the room, the evolution over time of the DBD actuator and the incoming velocity variation are all possible sources of fluctuations on the measured velocity.
However, the discrepancies are small and the overall performance of the control law is retained.

\begin{figure}
\centering
\begin{subfigure}{\textwidth}
  \centering
\includegraphics[]{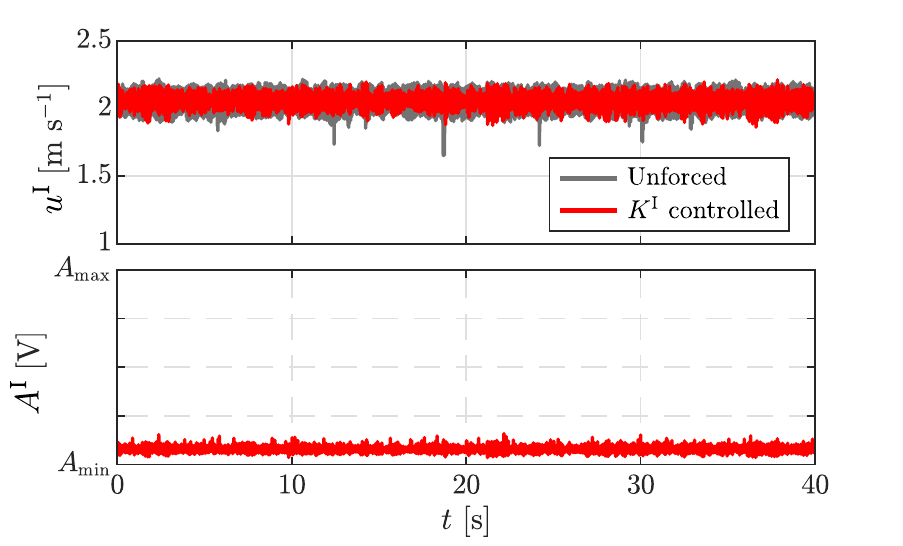}
\caption{}
\label{fig:R1_gMLC_TS}   
\end{subfigure}%

\begin{subfigure}{\textwidth}
  \centering
\includegraphics[]{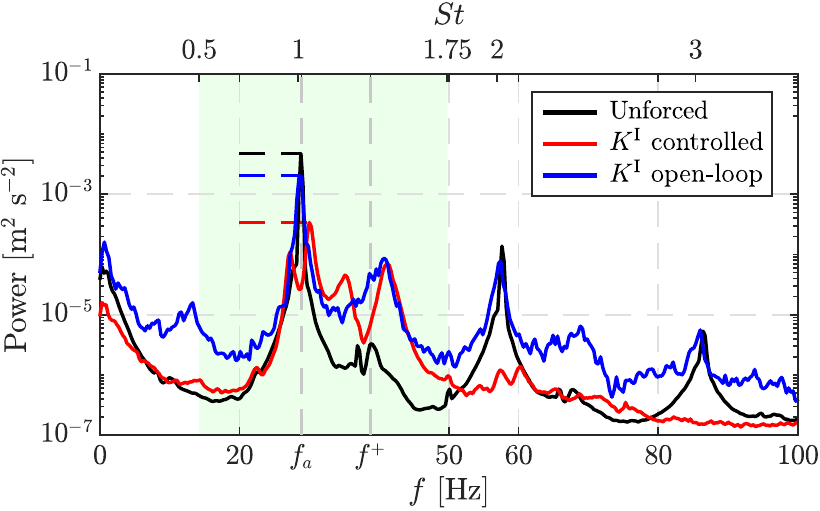}
\caption{}
\label{fig:R1_GMLC_Spectra}   
\end{subfigure}
\caption{\label{fig:Regime1_gMLC} Characteristics of the best control law $\KgMLCRa$ controlling the narrow-bandwidth regime.
(a) Time series of the velocity measured downstream and voltage level at the terminals of the DBD actuator.
(b) Power spectra of the velocity measured downstream.
Black is for the unforced flow, red for the flow controlled with $\KgMLCRa$ and blue for the flow controlled with the recorded actuation command $\ARa(t)$ employed as an open-loop control.
The spectra for the unforced flow (black) and the closed-loop controlled flow (red) are averaged over 20 realizations.
The horizontal dashed lines denote the maximum of the spectra in the observation window (green background).
The vertical axis is in $\log_{10}$ scale.
}
\end{figure}

Figure~\ref{fig:Regime1_gMLC} depicts the flow response under the control with the best control law $\KgMLCRa$ derived by gMLC for the narrow-bandwidth regime.
We observe that the control law $\KgMLCRa$ effectively answers to the control objective as the oscillation amplitude of the velocity measured for the controlled flow is reduced compared to the unforced flow, see figure~\ref{fig:R1_gMLC_TS}.
This goes along the decrease of the standard deviation to $\siggMLCRa = 61 \%$.
Such feature was consistently seen in different learning iterations with gMLC.
The power spectra on figure~\ref{fig:R1_GMLC_Spectra} show that $\KgMLCRa$ effectively reduces the highest peak of the spectrum at frequency $f_a$ by almost two orders of magnitude.
The effect of the control is also observed beyond the observation window between $\Str \in [0.5,1.75]$ as  the harmonics of $f_a$ are also nullified.
Note that the peak $f^+$ associated with the mode $n=3$ increases with the control.
This behavior is not surprising as the frequency is associated with a quasi-stable mode of the flow, which then grows when energy is supplied to the system.
Moreover, it appears that the frequency $f^+$ is split into two peaks.
This peak-splitting phenomenon, referred as \emph{spillover}, is well-known when building closed-loop transfer functions with unstable zeros and poles \citep{Rowley2006jfm,Rowley2006arfm}.
Despite being beyond a linear framework, one can suspect that the same mechanism is behind the observed peak splitting.
The learned law $\KgMLCRa$ is not able to control $f^+$ as its power level is around one order of magnitude lower than $f_a$ under control.

In addition, such control has been achieved with a minimum actuation power, indeed the associated cost is $\JbgMLCRa=0.0063$, less than $1\%$ of the maximum actuation power.
The actuation command, plotted on figure~\ref{fig:R1_gMLC_TS}, shows that the control is a combination of low level steady actuation  and a low amplitude feedback control.
To demonstrate the effectiveness of the low amplitude level, we use the closed-loop actuation command as an open-loop control.
Meaning that the closed-loop actuation command recorded during the learning process has been employed as open-loop control signal to force the flow.
The spectrum of the resulting flow (blue spectrum in figure~\ref{fig:R1_GMLC_Spectra}) shows that that the main frequency of the flow $f_a$ resurfaces and also that the mode $n=3$ associated to the frequency $f^+$ is also excited.
This open-loop test reveals that despite the low amplitude level, feedback plays a crucial role in stabilizing the flow.
Moreover, when we compare the results of the open-loop control with a steady actuation of equivalent level ($\sim 10\% A_{\rm max}$) in figure~\ref{fig:SAResponseR1}, we note that both frequencies $f_b$ and $f^+$ are excited, indicating the effect of the unsteady component of the command.

\begin{figure}
\centering
\includegraphics[]{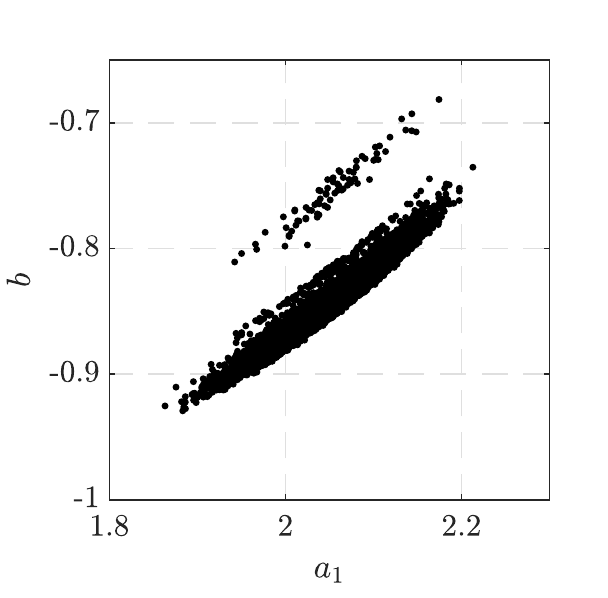}
\caption{Actuation command $b$ versus $a_1$ for the case: $\KgMLCRa$ controlling the narrow-bandwidth regime.}
\label{fig:K1a1}   
\end{figure}

Now we describe the learned law $\KgMLCRa$ with an analytical approximation and a cluster-based visualization.
For the analytical approximation, the determination coefficient for the affine reconstruction $R^2=0.87$ indicates an acceptable reconstruction.
The gains associated with each feature component are presented in table~\ref{tab:K1Coef}.
We note that, aside the mean component, the dominant term is $a_1=u$, as its gain  ($k_1$=0.51) is more than 7 times higher than the second highest gain.
The strong correlation between $\KgMLCRa$ and $a_1$ reveals that phasor control or direct feedback of the system's state plays a major role for this control.
However, figure~\ref{fig:K1a1} shows that the relationship between $b$ and $a_1$ is not fully affine as two regions of significant width are displayed.
This analysis is in agreement with the cluster-based investigation of the control law.
\begin{table}
\renewcommand{\arraystretch}{1.25}
\setlength{\tabcolsep}{6pt}
\begin{center}
\begin{tabular}{lcccccccccccc}
Term & 1 & $a_1$ & $a_2$ & $a_3$ & $a_4$ & $a_5$ & $a_6$ & $a_7$ & $a_8$ & $a_9$ & $a_{10}$ \\
Gain & $k_0$ & $k_1$ & $k_2$ & $k_3$ & $k_4$ & $k_5$ & $k_6$ & $k_7$ & $k_8$ & $k_9$ & $k_{10}$ \\
Value & -1.50  & 0.51  & -0.06 & -0.07 & 0.01 & -0.06 & 0.02 & -0.04 & 0 & 0 & 0
\end{tabular} 
\end{center}
\caption{Gains for the affine reconstruction of $\KgMLCRa$ controlling the narrow-bandwidth regime.}
\label{tab:K1Coef}
\end{table}

\begin{figure}
\centering
\begin{subfigure}{.44\textwidth}
  \centering
\includegraphics[height=6cm]{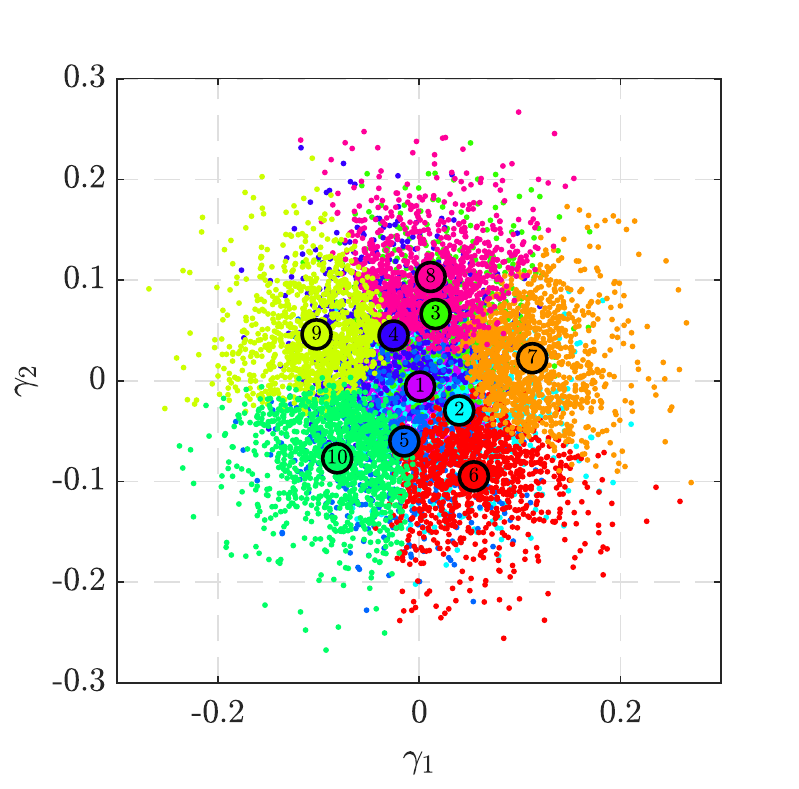}
\caption{Proximity map of the feature vector $\bm a$.}
\label{Fig:PM_K1}   
\end{subfigure}%
\hfil
\begin{subfigure}{.48\textwidth}
  \centering
\includegraphics[height=6cm]{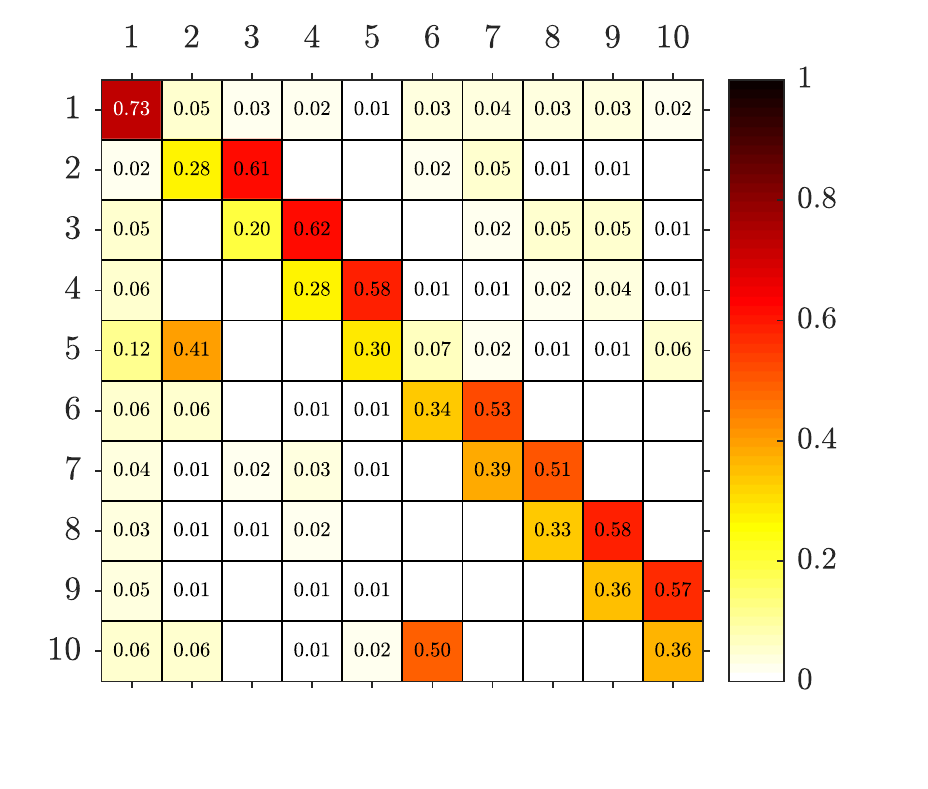}
\caption{Transition probability matrix.}
\label{Fig:PM_TrMat_K1}   
\end{subfigure}

\begin{subfigure}{.5\textwidth}
  \centering
\includegraphics[width=\linewidth]{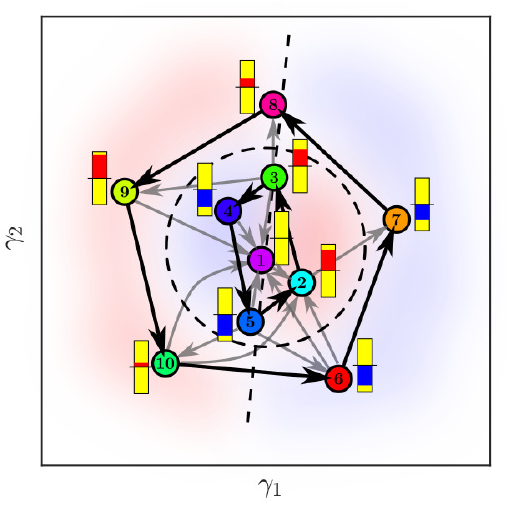}
\caption{Control network.}
\label{Fig:ControlNet_K1}   
\end{subfigure}%
\caption{\label{Fig:GraphInterpretation_1} Investigation of the law $\KgMLCRa$ controlling the narrow-bandwidth regime.
Clusters are painted in different colors.
The centroids are marked with numbers from 1 to 10.
Sub-figure (c) represents the control network model.
The actuation amplitudes are denoted with bars, red (blue) for positive (negative) amplitude with respect to the mean actuation.
The yellow boxes underneath indicate 25\% of the maximum amplitude.
The black arrows serve as the most probable transition from one cluster to the other ($p>0.40$).
The gray arrows are for less probable transitions ($0.40 \geq p \geq 0.05$).
Lower probability transitions and self transitions are omitted for clarity.
The red (blue) background denotes the supposed regions of positive (negative) amplitude.
The dotted black lines are the supposed limit between the regions.
}
\end{figure}

The control visualization, following the method described in \S~\ref{Sec:ControlLawInterpretation}, makes it possible to better specify the dynamics of the control achieved by $\KgMLCRa$.
The proximity map (figure~\ref{Fig:PM_K1}) shows that the centroids are arranged in a circular manner around the origin where the centroid 1 is located.
The transition probability matrix (figure~\ref{Fig:PM_TrMat_K1}) gives the probability transition from one cluster to the other for each time step $\rm d t=\SI{0.004}{\second}$.
The transition probability matrix also shows one fixed point and 2 cycles: one small with centroids 2, 3, 4 and 5 on one hand and one large with centroids 6, 7, 8, 9 and 10 on the other hand.
Combining the proximity map and the transition probability matrix, we can reconstruct the phase space of the dynamics by deriving a control network model (figure~\ref{Fig:ControlNet_K1}).
The identified cycles in the transition probability matrix are then represented by  limit cycles in the phase space.
A frequency analysis based on a Poincar\'e section and angular first return map of the dynamics similar to \citet{Lusseyran2008pof} reveals that the frequencies associated to the large limit and small limit cycles are around $\SI{28.48}{\hertz}$ and $\SI{41.14}{\hertz}$ respectively.
We can then assume that the large limit cycle is associated with the dynamics of mode $n=2$ (frequency $f_a=28.81$) and the small limit cycle is associated with mode $n=3$ (frequency $f^+=38.72$).
Following \citet{Lusseyran2008pof}, such organization frequencies in the phase space shows that the dynamics may be structured around a fixed point of the stable spiral type where the oscillation's time period increases with the distance to the fixed point, here represented by cluster 1.
Figure~\ref{Fig:ControlNet_K1} also depicts the actuation regions in the phase space, revealing two mains regions of opposite actuation sign separated by a straight line.
Interestingly, the sign of the actuation changes when approaching the fixed point.
Such actuation map shows that the main stabilization mechanism exploited by $\KgMLCRa$ is phasor control.
The change of actuation sign when approaching the fixed point is explained by a phase shift due to the frequency change. 
The resulting control is similar to the one obtained with linear control by \citet{Yan2006aiaa} where there is a rapid switching between two modes competing for the available energy and thus mitigating any resonance.
Moreover, a spectral analysis of the actuation command $b$ and $a_1=u$ shows that they share the same frequency peaks comforting the phase relation between the control and the state of the system, see figure~\ref{fig:R1K1} in appendix~\ref{appD}.

Finally, a comparison between MLC and gMLC has been performed (see appendix~\ref{appB}) and reveals that gMLC outperforms MLC in terms of speed and final solution.
In total, the learning has been accelerated by one order of magnitude.




Gradient-enriched MLC has been successfully applied to the stabilization of the open-cavity flow experiment.
Exploration and exploitation phases both participated to the fast learning of a feedback control law.
The evolution phases managed to discover new minima in the search space and the simplex steps succeeded in converging towards a new minimum.
A feedback control law is built, outperforming the steady actuation and allowing a similar reduction of the level of the maximum peak of the spectrum but with small actuation power.
Both the analytical approximation and the cluster-based analysis hints a control combining  phasor control and nonlinear interactions.
Hence, the control achieved is close to an ideal stabilization scenario, where some kind of base state or fixed point, is stabilized with a vanishing cost.
This interpretation is certainly to be considered heuristically given the complexity of the real dynamics of the 3D intra-cavity flow and its nonlinear temporal and spatial interactions with the mixing layer.
However, it aims at capturing the remarkable properties of the control law learned by the gMLC, whose mode of action is radically opposed to that of a control by steady forcing.
In this section, the learning has been done for a single frequency regime, in the next section, a more challenging regime is controlled where two modes compete, strengthening nonlinear coupling.

\subsection{Closed-loop control of the mode-switching regime}\label{sec:gmlc_R2}
In this section, gMLC is employed to control a flow regime with strong nonlinear coupling which lead, in particular, to an intermittency between the two main instability modes of the mixing layer.
The dynamics of intermittency is chaotic \citep{Lusseyran2008pof} with the appearance of long time scales that are demanding from the point of view of machine learning.
The goal is again to stabilize the flow by reducing the oscillations of the mixing layer but this time in the case where two modes compete as described in \S~\ref{Sec:Unforced_dynamics}.
This constitutes a challenging problem as gMLC needs to learn a control law able to control two modes simultaneously.
For the gMLC optimization, the same parameters as for the narrow-bandwidth regime have been employed, see \S~\ref{sec:gmlc_settings}.
In the following, the notations for the control law, cost and standard deviation associated with the learning on the mode-switching regime are marked by the superscript $\textrm{II}$.

\begin{figure}
\centerline{\includegraphics[width=0.90\linewidth]{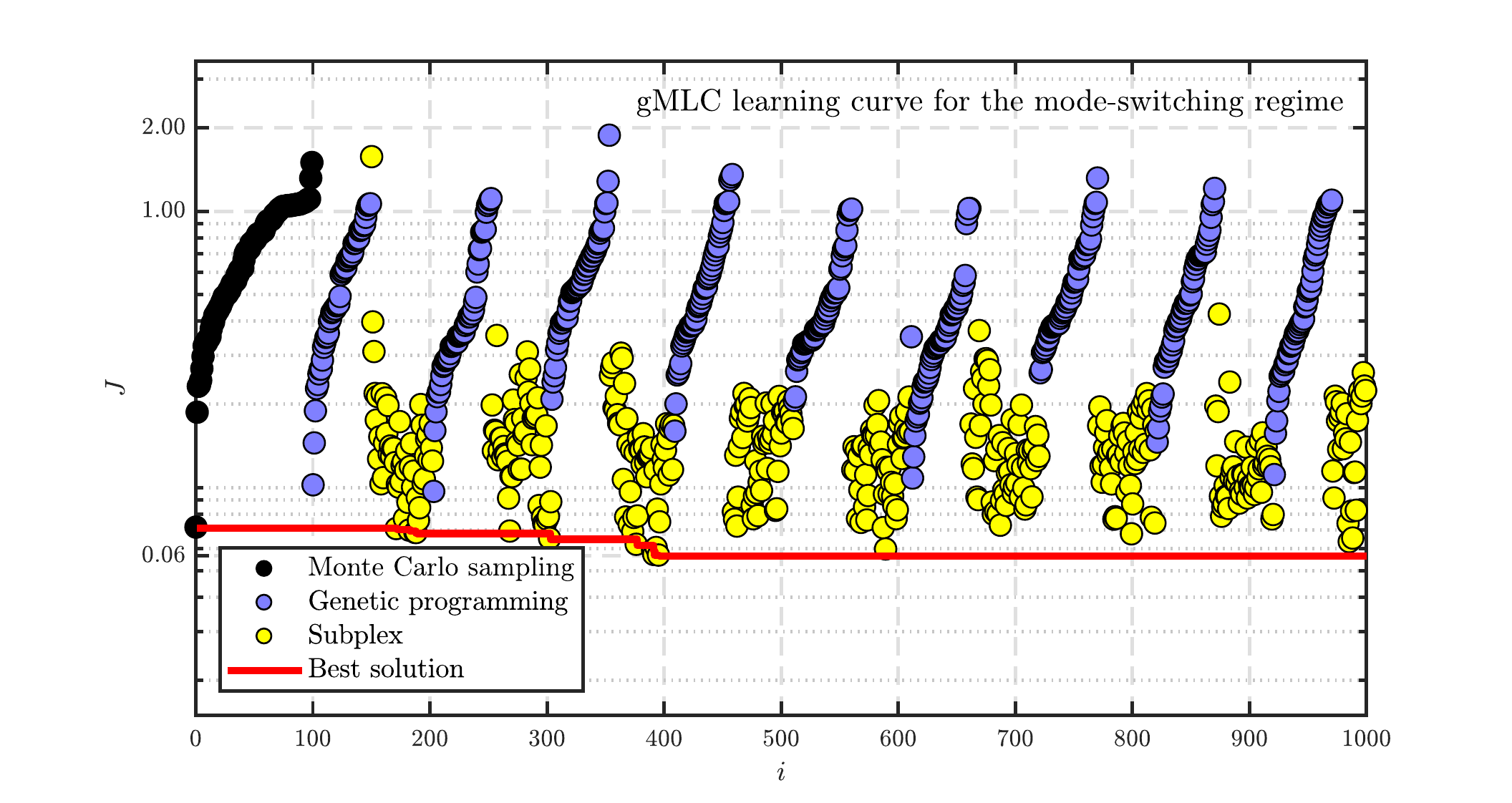}}
 \caption{Distribution of the costs for the 1000 evaluated individuals during the gMLC optimization process for the mode-switching regime.
Each dot represents the cost $J$ of one individual.
The color of the dots symbolize how the individuals have been generated: 
black dots for randomly generated individuals (Monte Carlo sampling phase), blue dots for individuals generated from a genetic operator (exploration phase) and yellow dots for the the individuals arising from the subplex method (exploitation phase).
The individuals from the Monte Carlo sampling and exploration phases are sorted following their costs.
The red line follows the evolution of the best cost.
The vertical axis is in $\log_{10}$ scale.}
\label{fig:ProgressR2}
\end{figure}

Figure~\ref{fig:ProgressR2} depicts the cost of the individuals evaluated along the learning process.
For this specific experiment, most of the learning is realized during the Monte Carlo sampling phase, reducing the cost function to $J=0.0713$.
From there on, the only improvements are carried out with the simplex steps, bringing the cost function to the final value $\JgMLCRb=0.0565$.
Nonetheless, the second exploration phase introduced a new control law ($\#11$) in the simplex (see table~\ref{tab:K_gMLC_R2} of appendix~\ref{appC}).
As the gMLC algorithm is partially stochastic, it is possible to fall close to the global minimum by pure luck but it usually takes several iterations of the learning phases to converge, as in \S~\ref{sec:gmlc_R1}.
Such learning process has been observed in other realizations of the same experiment where a combination of exploration and exploitation have been necessary to reach similar levels of performance.

\begin{figure}
\centering
\begin{subfigure}{\textwidth}
  \centering
\includegraphics[width=0.75\linewidth]{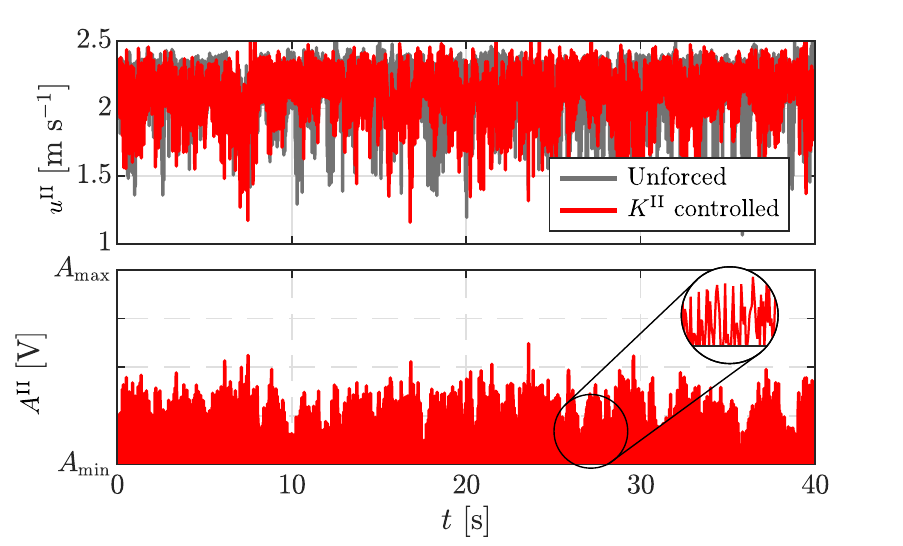}
\caption{}
\label{fig:R2_gMLC_TS}   
\end{subfigure}%

\begin{subfigure}{\textwidth}
  \centering
\includegraphics[width=0.75\linewidth]{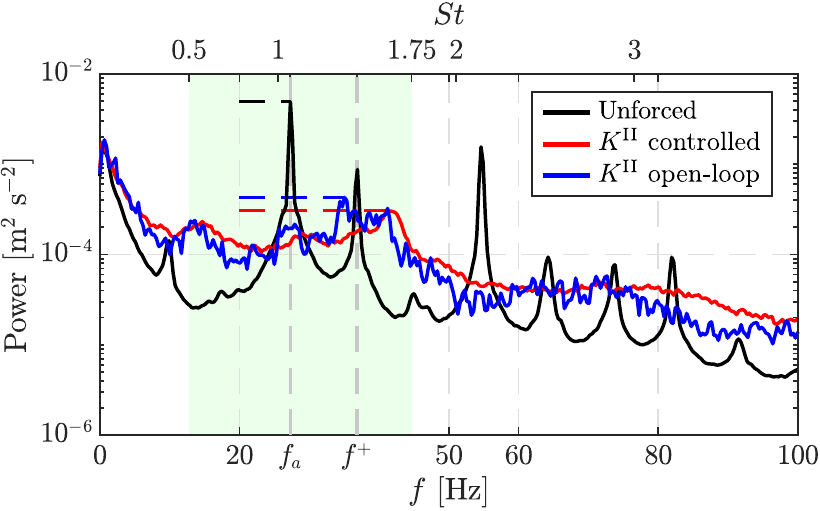}
\caption{}
\label{fig:R2_GMLC_Spectra}   
\end{subfigure}
\caption{\label{fig:Regime2_gMLC} Characteristics of the best control law $\KgMLCRb$ controlling the mode-switching regime.
(a) Time series of the velocity measured downstream and voltage level at the terminals of the DBD actuator.
A short window is enlarged to display the behavior close to $A_{\rm min}$ (the proportions are not kept).
(b) Power spectra of the velocity measured downstream.
Black is for the unforced flow, red for the flow controlled with $\KgMLCRb$ and blue for the flow controlled with the recorded actuation command $\ARb(t)$ employed as an open-loop control.
The spectra for the unforced flow (black) and the closed-loop controlled flow (red) are averaged over 20 realizations.
The horizontal dashed lines denote the maximum of the spectra in the observation window (green background).
The vertical axis is in $\log_{10}$ scale.
}
\end{figure}

\begin{figure}
\centering
\includegraphics[width=0.5\textwidth]{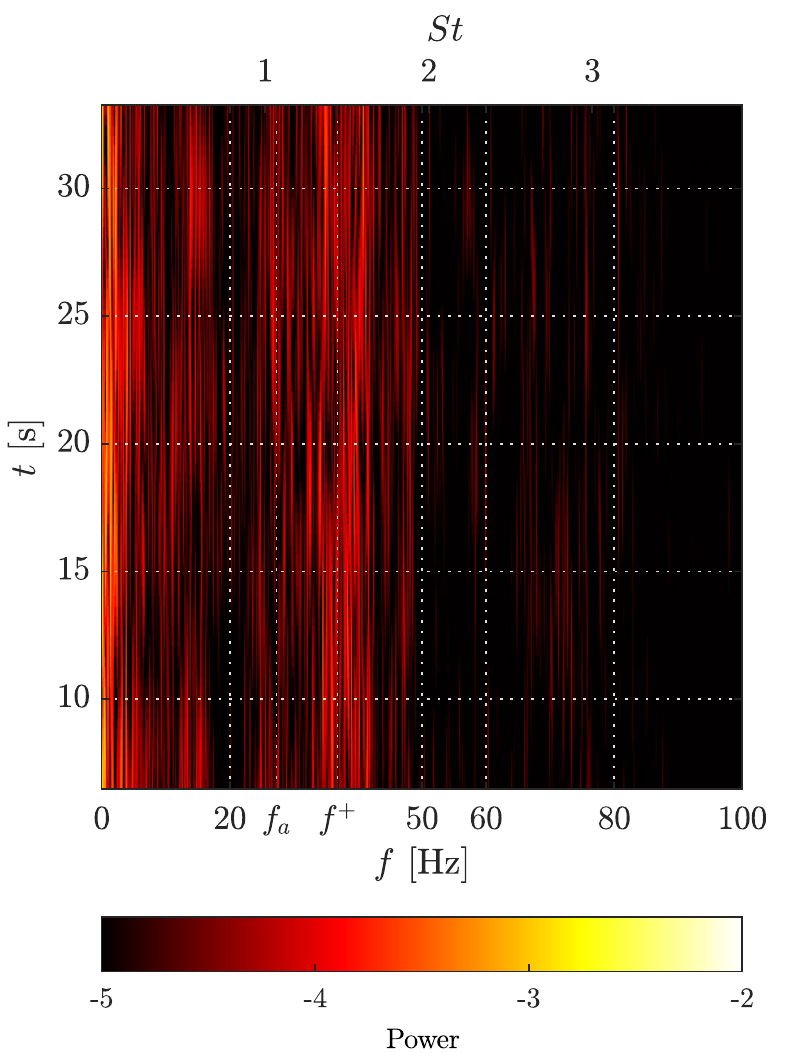}
\caption{\label{fig:Spectrocontrolled_R2}Spectrogram of the velocity for the mode-switching regime controlled with $\KgMLCRb$.}
\end{figure}

After 1000 evaluations, the final control $\KgMLCRb$ is a feedback control law, thoroughly described in appendix~\ref{appC}.
The spectra of the flow under control (figure~\ref{fig:R2_GMLC_Spectra}) reveals a drastic decrease in the level of the maximum peak at frequency $f_a$.
The dominant frequency is then close to the one associated with the second main mode ($f^+$).
The relative reduction of the maximum peak in the spectrum is $\JagMLCRb=0.0335$, i.e. 0.97\% of the fluctuation level.
This corresponds to a reduction of the oscillation's amplitude by a factor 5 or by $\SI{15}{\decibel}$.
The standard deviation associated decreases to $\siggMLCRb = 97 \%$.
Like for the narrow-bandwidth regime, the control is achieved with small actuation power, using around $2\%$ of the maximum actuation power.
Figure~\ref{fig:Spectrocontrolled_R2} shows that the controlled flow does no longer shows a mode-switching behavior like in figure~\ref{fig:SpectrogramIntermittent}. 
The actuation command plotted in figure~\ref{fig:R2_gMLC_TS} shows a short-time intermittent high-amplitude spikes emerging from the minimum actuation level ($A_{\rm min}$).
Like for the narrow-bandwidth regime, the control law $\KgMLCRb$ is re-evaluated 20 times and
a small performance drop is observed: $\JgMLCRb$ went from $-94\%$ to $-89\%$ and the standard deviation from $\siggMLCRb=97\%$ to $\widetilde{\sigma}=112\%$.
Finally, like in \S~\ref{sec:gmlc_R1}, the time series of the actuation command has been employed as a signal input for an open-loop control.
Surprisingly, the equivalent open-loop control performs as good as the close-loop control law, suggesting that feedback was not at play in the reduction of the power amplitude for the mode-switching regime.
However, anticipating on the next section (\S~\ref{Sec:Discussion}), controlling the narrow-bandwidth regime with $\KgMLCRb$ reveals that feedback is still a feature selected by gMLC.

For the polynomial approximation of the $\KgMLCRb$ control of the mode-switching regime, a linear regression was unable to derive an affine reconstruction of the actuation command; the corresponding determination coefficient is $R^2=0.13$.
Even with expanding the affine reconstruction with quadratic and cubic terms, $R^2$ is less than 0.25, implying that no linear control can be inferred from the data.

\begin{figure}
  \centering
\includegraphics[width=0.75\linewidth]{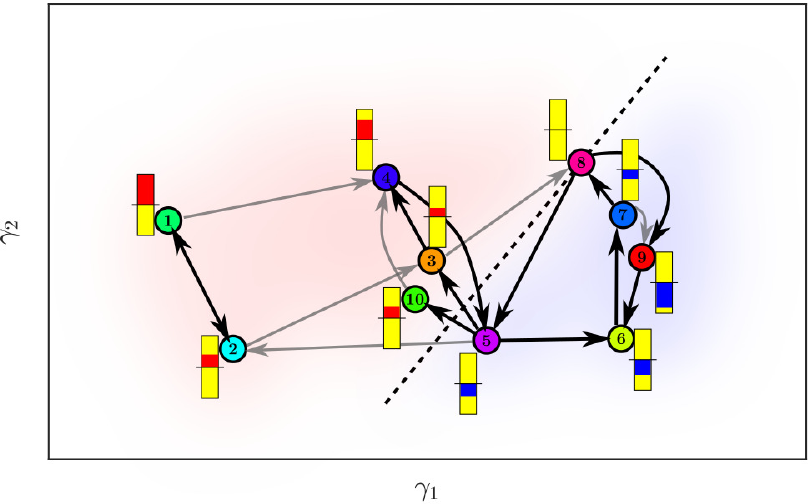}
\caption{\label{Fig:GraphInterpretation_2} Visualization of the control law $\KgMLCRb$ controlling the mode-switching regime.
The control network is depicted as in figure~\ref{Fig:GraphInterpretation_1}.
The actuation amplitudes are denoted with bars, red (blue) for positive (negative) amplitude with respect to the mean actuation.
The yellow boxes indicate 10\% of the maximum amplitude.
The black arrows serve as the most probable transition from one cluster to the other ($p>0.15$).
The gray arrows are for are for less probable transitions ($0.15 \geq p \geq 0.10$).
Lower probability transitions and self transitions are omitted for clarity.
The red (blue) background denotes the supposed regions of positive (negative) amplitude.
The dotted black lines are the supposed limit that separates the regions.
}
\end{figure}

Regarding the cluster-based analysis of the control, figure~\ref{Fig:GraphInterpretation_2} reveals a complex structure for the dynamics.
The transition matrix (not plotted) shows that the cluster self-transitions are dominant; they are not displayed in the control network for clarity.
The reconstructed phase space reveals two regions of opposite actuation signs.
The complex interactions between the centroids shows that strong nonlinearities are at play.
Interestingly, the spectrum of the actuation command (figure~\ref{fig:R2K2} in appendix~\ref{appD}) does not include any significant peak, except for the very low frequencies.
The actuation command corresponds then to a random noise without any correlation with the measured velocity.

In less than 1000 evaluations, gMLC manage to build feedback control laws that reduced the maximum peak of the power spectrum with small actuation power in two regimes: the narrow-bandwidth regime and the mode-switching regime.
We proposed a visual representation of the control laws to aid the interpretability of the actuation mechanisms that enabled such efficient controls.
However, a real analysis of the controlled flow is not done yet and constitutes a study in itself.
The identification of the involved control mechanism requires a study of the short transient that leads to the stabilized state.
Nonetheless, we can affirm that the control impacts the linear amplifier of the shear layer as the DBD actuator modifies its thickness on average.
This effect has been demonstrated by studying the difference between the unforced and forced mean flow even for low actuation levels, i.e., near the ionization threshold.
Such mechanism is expected to remain valid for turbulent flows and even for often studied transonic cavity flows.
Moreover, \citet{CornejoMaceda2021jfm} show that gMLC surpasses MLC in terms of performance of the final solution and learning speed for a 2D numerical simulation.
In appendix~\ref{appA}, we demonstrated that gMLC also surpasses MLC in experiments, establishing gMLC as a keystone for fast learning of feedback control laws directly on the plant.
We foresee that gMLC will greatly contribute the learning of control laws for MIMO control.

%% file: S5.tex
\section{Control law investigation}
\label{Sec:Discussion}
In this section, we further investigate the capabilities of the control laws $\KgMLCRa$ and $\KgMLCRb$ learned for the narrow-bandwidth regime and the mode-switching regime respectively.
Firstly, the robustness of the laws is tested by applying each law on the other regime, comprising dynamics different from the learning conditions (\S~\ref{Sec:InterpCross}).
Secondly, we characterize the new control of these regimes with an affine approximation and our cluster-based visualization method (\ref{Sec:InterpCross}).
Finally, we establish the existence of an effective feedback for the control of both regimes with open-loop tests (\S~\ref{sec:discuss_feedback}).


\subsection{Robustness of the control}\label{sec:discuss_robustness}
In this study, feedback control laws are optimized for steady conditions: fixed Reynolds number and incoming velocity during the learning process.
To achieve robustness, the control laws learned should also perform for a large range of parameters.
Thus, to test the robustness of the learned laws, they are re-evaluated for operating conditions different from the learning ones: the law $\KgMLCRa$ optimized for the narrow-bandwidth regime is now employed to control the mode-switching regime and vice versa.
Such test is demanding as the dynamics are partially different from one regime to the other.
In particular, the unforced dynamics of the mode-switching regime includes the main frequency of the narrow-bandwidth regime but also display an intermittency with another frequency.

\begin{figure}
\centering
\begin{subfigure}{\textwidth}
  \centering
\includegraphics[]{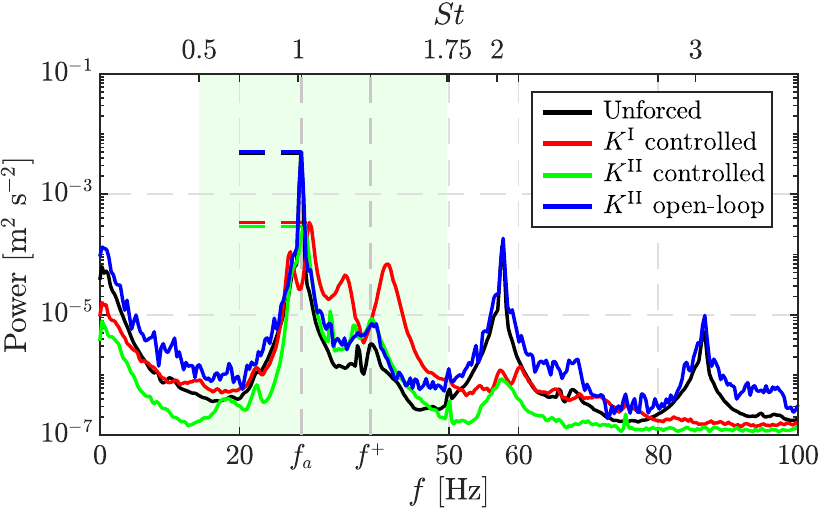}
\caption{Power spectra for the narrow-bandwidth regime.}
\label{fig:K2onR1}   
\end{subfigure}%

\begin{subfigure}{\textwidth}
  \centering
\includegraphics[]{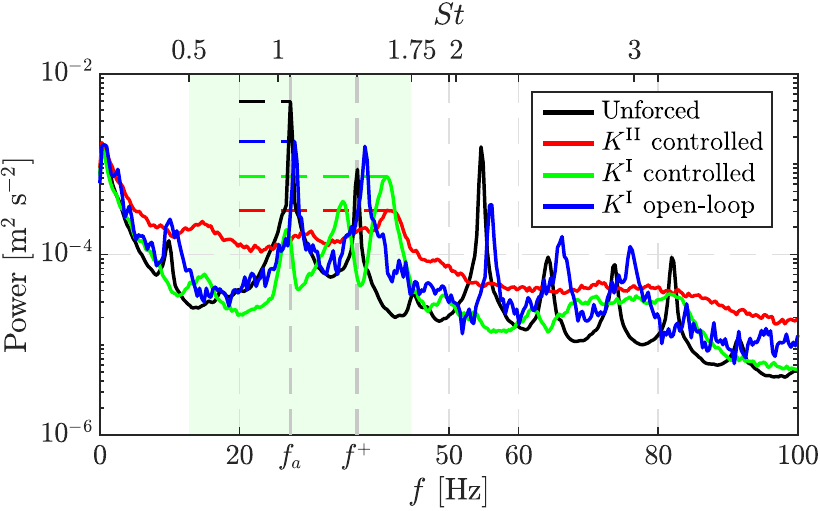}
\caption{Power spectra for the mode-switching regime.}
\label{fig:K1onR2}   
\end{subfigure}
\caption{\label{fig:ControlOtherRegime}
Spectral response of the flow controlled by $\KgMLCRa$ and $\KgMLCRb$ for the narrow-bandwidth regime (top) and the mode-switching regime (bottom).
Black spectra corresponds to the unforced dynamics of each regime;
Red spectra corresponds to the flow controlled by law learned in the given regime, i.e., $\KgMLCRa$ ($\KgMLCRb$) for the narrow-bandwidth (mode-switching) regime.
Green corresponds to the flow controlled by the law learned in the other regime,
and blue to the open-loop equivalent of the latter one.
The horizontal dashed lines denote the maximum of each spectrum in the observation window (shaded green section).
The vertical axis are in $\log_{10}$ scale.
}
\end{figure}

Figure~\ref{fig:ControlOtherRegime} displays the spectra for each controlled case.
For the control of the narrow-bandwidth regime with $\KgMLCRa$ (figure~\ref{fig:K2onR1}, green line), we note that the law $\KgMLCRb$ manages to reduce the main peak $f_a$ of the spectrum to the same level as the law $\KgMLCRa$ (red line).
Moreover, we can notice that for the frequency range around $f^+$, $\KgMLCRb$ performs better than $\KgMLCRa$ where it manages to nullify the peak-splitting phenomenon described in \S~\ref{sec:gmlc_R1}.
Such feature is expected as $\KgMLCRb$ has been optimized to also control the mode $f^+$.
In fact, the whole spectral range is reduced in amplitude.
This non-transfer of energy to other frequencies is a remarkable feature of $\KgMLCRb$.

As for the control of the mode-switching regime with $\KgMLCRa$ (figure~\ref{fig:K1onR2}, green line), the power of the main frequency $f_a$ is drastically reduced, reaching the same power level as the control with $\KgMLCRb$ (red line).
However, the controller is less efficient than $\KgMLCRb$, as $\KgMLCRa$ fails to reduce the power associated to the frequency range surrounding the frequency $f^+$ of mode $n=3$.
Again, like for the control of the narrow-bandwidth regime with $\KgMLCRa$, we observe the spillover effect with a splitting of the main frequency into two frequencies on either side and with a lower power level.

In summary, both learned control laws $\KgMLCRa$ and $\KgMLCRb$ are able to retain their efficiency when controlling regimes that are out of the learning conditions.
Expectedly, $\KgMLCRa$ was unable to reduce the peak of the frequency $f^+$ in the mode-switching regime but it manages to reduce the $f_a$ peak as it was built for, despite appearing only intermittently.
On the other hand, for the narrow-bandwidth regime, the control with $\KgMLCRb$ was more significant than $\KgMLCRa$ as it manage to also control prevent the peak-splitting (or spillover) of the third mode.
Thus this test also reveals, that $\KgMLCRb$ is not only able to control the frequency $f_a$ and $f^+$ but also to prevent the rise of both modes simultaneously.
Of course, from the point of view of the cost function, both control laws $\KgMLCRa$ and $\KgMLCRb$ are similar when controlling the narrow-bandwidth regime and gMLC could have converges towards any of the two control laws.
Yet, $\KgMLCRb$ is the control law that answers the best the control objective which is the stabilization of the flow.
Therefore, it is really the learning conditions that make the difference.
This analysis reveals that learning a control in complex and rich conditions is beneficial for the robustness and for the overall efficiency of the control as the richness of the dynamics will be reflected in the control law.

\subsection{Interpretation of the resulting controlled flow}\label{Sec:InterpCross}
Now, we propose an interpretation of the controls performed in the previous section using both analytical approximation and cluster-based visualization of the control laws.
Firstly, we analyze the case where $\KgMLCRb$ controls the narrow-bandwidth regime.
Like for \S~\ref{sec:gmlc_R2}, a linear regression is unable to reconstruct the actuation command despite being in a less complex regime.
The determination coefficient is $R^2=0.13$.
The addition of quadratic terms brings the coefficient no higher than 0.78.
A complex nonlinear control is expected as $\KgMLCRb$ manages to control both frequencies $f_a$ and $f^+$.

\begin{figure}
 \centering
\includegraphics[width=0.55\linewidth]{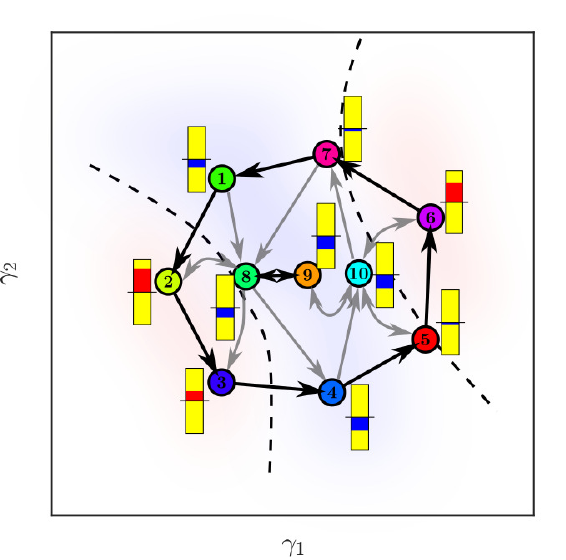}
\caption{\label{Fig:GraphInterpretation_3} Visualization of the control law $\KgMLCRb$ controlling the narrow-bandwidth regime.
The control network is depicted as in figure~\ref{Fig:GraphInterpretation_1}.
The actuation amplitudes are denoted with bars, red (blue) for positive (negative) amplitude with respect to the mean actuation.
The yellow boxes indicate 25\% of the maximum amplitude.
The black arrows serve as the most probable transition from one cluster to the other ($p>0.50$).
The gray arrows are for less probable transitions ($0.50 \geq p \geq 0.5$).
Lower probability transitions and self transitions are omitted for clarity.
The red (blue) background denotes the supposed regions of positive (negative) amplitude.
The dotted black lines are the supposed limit that separates the regions.
}
\end{figure}

As for the control law visualization, figure~\ref{Fig:GraphInterpretation_3} depicts a similar control network as in \S~\ref{sec:gmlc_R1}.
However, there is only one large cycle composed of the centroids 1, 2, 3, 4, 5, 6 and 7.
The limit cycle presents four phases regularly alternating between positive and negative actuation,
suggesting that the control operates at twice the frequency of the flow.
A spectral analysis of $a_1=u$ and $b$ (see figure~\ref{fig:R1K2} in appendix~\ref{appD}) shows that the main peaks are respectively $f_a=\SI{29.03}{\hertz}$ and $f=\SI{58.23}{\hertz} \approx 2f_a$, confirming the phase relation between the flow dynamics and the actuation command.
Interestingly, the second harmonic $2f_a$ is slightly excited but does not resonate for the controlled flow (figure~\ref{fig:K2onR1}, green line), 
while it clearly resonates for the open-loop equivalent of $\KgMLCRb$ (blue line).
In summary, $\KgMLCRb$ controls the flow at twice the main frequency but avoids the resonance of the second harmonic.
The efficiency of a control at twice the main frequency has been previously reported by \citet{Schumm1994jfm}.
The authors note the stabilization of a cylinder wake by transverse vibration of the cylinder at 1.8 times the natural shedding frequency.
In particular, they declare that the control effect is due to a nonlinear interaction between the instability and the forcing input.
Thus, both, the analytical and cluster-based analysis point towards a nonlinear actuation mechanism for the control of the narrow-bandwidth with $\KgMLCRb$.

%
%
   
\begin{figure}
\centering
\includegraphics[]{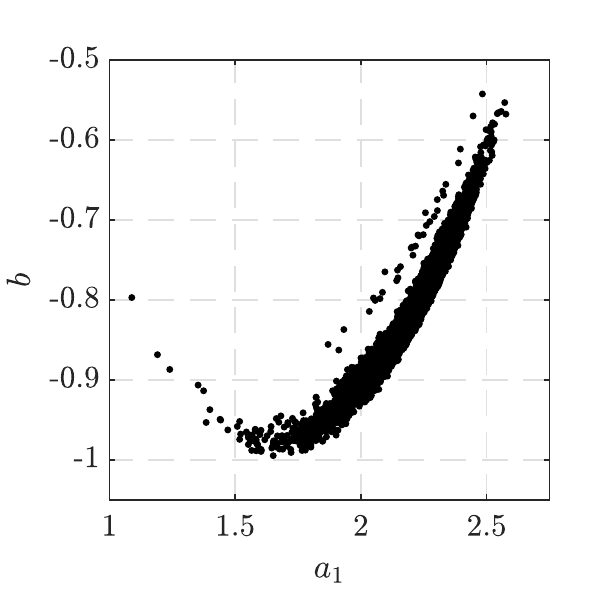}
\caption{Actuation command $b$ versus $a_1$ for the case: $\KgMLCRa$ controlling the mode-switching regime.}
\label{fig:K4a1}   
\end{figure}

\begin{figure}
  \centering
\includegraphics[width=0.55\linewidth]{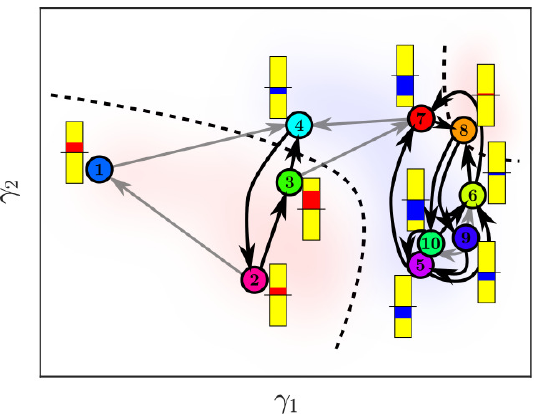}
\caption{\label{Fig:GraphInterpretation_4} Visualization of the control law $\KgMLCRa$ controlling the mode-switching regime.
The control network is depicted as in figure~\ref{Fig:GraphInterpretation_1}
The actuation amplitudes are denoted with bars, red (blue) for positive (negative) amplitude with respect to the mean actuation.
The yellow boxes indicate 25\% of the maximum amplitude.
The black arrows serve as the most probable transition from one cluster to the other ($p>0.15$).
The gray arrows are for less probable transitions ($0.15 \geq p \geq 0.9$).
Lower probability transitions and self transitions are omitted for clarity.
The red (blue) background denotes the supposed regions of positive (negative) amplitude.
The dotted black lines are the supposed limit that separates the regions.
}
\end{figure}

Secondly, we interpret the case where $\KgMLCRa$ controls the mode-switching regime.
This time, the linear regression manages to build an affine approximation of the control as the determination coefficient is $R^2=0.92$.
The gains associated with each feature component are displayed in table~\ref{tab:K2Coef}.
Like for \S~\ref{sec:gmlc_R1}, the most relevant feature is $a_1=u(t)$ but again the control cannot be reduced to an affine relationship as figure~\ref{fig:K4a1} displays a nonlinear curve.
Such observation is in agreement with the spectral analysis of $b$ and $a_1=u$ (see figure~\ref{fig:R2K1} in appendix~\ref{appD}) showing the peaks at the same frequency.
The complexity of the flow translates into a complex control network as in \S~\ref{sec:gmlc_R2}.
Figure~\ref{Fig:GraphInterpretation_4} depicts a reconstructed phase space divided into two main regions: one on the left (centroids 1, 2 and 3) with positive actuation amplitude; and one on the right (centroids 4, 5, 6, 7, 9 and 10) with negative actuation amplitude.
The role of centroid 8 may be small as its associated actuation is close to the mean value.
Interestingly, the overall structure of the control network is similar to the one on figure~\ref{Fig:GraphInterpretation_4}, suggesting that the control mechanism is also a complex nonlinear one.

\begin{table}
\renewcommand{\arraystretch}{1.25}
\setlength{\tabcolsep}{6pt}
\begin{center}
\begin{tabular}{lcccccccccccc}
Term & 1 & $a_1$ & $a_2$ & $a_3$ & $a_4$ & $a_5$ & $a_6$ & $a_7$ & $a_8$ & $a_9$ & $a_10$ \\
Gain & $k_0$ & $k_1$ & $k_2$ & $k_3$ & $k_4$ & $k_5$ & $k_6$ & $k_7$ & $k_8$ & $k_9$ & $k_{10}$ \\
Value & -1.65  & 0.48 & -0.06 & -0.05 & 0.01 & -0.01 & 0 & 0 & 0 & 0.01 & 0
\end{tabular} 
\end{center}
\caption{Gains for the affine reconstruction of $\KgMLCRa$ controlling the mode-switching regime.}
\label{tab:K2Coef}
\end{table}

\subsection{The need of feedback}\label{sec:discuss_feedback}
We've shown on the control of the narrow-bandwidth regime (\S~\ref{sec:gmlc_R1}) that without feedback the control law $\KgMLCRa$ is unable to stabilize the flow and even excites the frequencies $f_a$ and $f^+$ (figure~\ref{fig:R1_GMLC_Spectra}).
Moreover, $\KgMLCRa$ is able to partially control the mode-switching regime (\S~\ref{sec:discuss_robustness}) and, as expected, the same controller applied in an open-loop manner is no longer able to control the main mode $f_a$ (blue spectrum in figure~\ref{fig:K1onR2}).
We note, nonetheless, a small shift of the spectrum towards the higher frequencies.

On the other hand, surprisingly, it has been shown in (\S~\ref{sec:gmlc_R2}) that $\KgMLCRb$ performs in open-loop as well as in closed-loop.
This result is consistent with the absence of correlation between the actuation command $b$ and the velocity measure $a_1=u$, see figure~\ref{fig:R2K2} in appendix~\ref{appD}.
So, it seems that the learning process has selected the flow information ($a_i$, see appendix~\ref{appC}) without any improvement of the cost.
However, $\KgMLCRb$ applied in closed-loop manner to the narrow-bandwidth regime performs even better than $\KgMLCRa$, while in an open-loop manner, $\KgMLCRb$ fails to achieve any control.
Indeed, the corresponding spectrum (blue spectrum in figure~\ref{fig:K2onR1}) is almost similar to the unforced flow.
Therefore, it should be noticed that the flow state information $a_i$ in $\KgMLCRb$ are truly functional and in fact give the controller the ability to remain effective well away from the learning conditions.

This analysis shows the extent to which feedback is a key feature for control.
We believe that the ability of gMLC to learn effective and efficient feedback control laws in experiments will greatly benefit future MIMO control experiments.

\subsection{Stabilization of the open cavity flow}
In this study, the flow is monitored by a single hot-wire sensor downstream of the actuator.
The sensor signal is employed for sensor feedback and to characterize the controlled flow.
The achieved stabilization near the sensor extends in spanwise and streamwise direction,
assumingly a significant portion of the finite aspect ratio cavity.

For large aspect ratios $S/D$, e.g.\ $O(100)$ or more, the effect of two-dimensional actuation
along the whole span can be expected to depend on the sensor location.
The feedback stabilization in the sensor plane will become a non-stabilizing open-loop actuation
far beyond the spanwise coherence length.
An interesting example has been reported for the stabilization of a large-aspect-ratio cylinder wake.
\citet{Roussopoulos1993jfm} 
forced the cylinder wake with a pair of loudspeakers driven in opposite phase
and significantly reduced the fluctuations at the downstream sensor location.
Far away from the sensor in spanwise direction, 
no stabilization was observed.
For our open cavity flow, we expect a loss of control authority at a distance from the hot-wire position greater than the transverse coherence length.
This transverse coherence of the mixing layer instabilities is at least of the order of the cavity depth $D$.
Our cavity has an aspect ratio of $S/D$=6, i.e. we control at least one third of the cavity spanwise 
(1D on either side of the hot-wire plane).
The remaining lateral thirds are in the Ekman layers and therefore probably less oscillating.
The global stabilization could be augmented with multiple actuators and multiple sensors.

As for the spanwise homogeneity of our one-piece DBD actuator, measurements performed for actuation levels close to the ionization threshold, when the ionization is still quite inhomogeneous along the electrode, show that the flow response is independent of the spanwise location of the measurement point.

Concerning the streamwise direction, the source of the oscillation is related to the mixing layer through the Kelvin-Helmholtz instability.
At the level of the mixing layer, the hypothesis of an oscillation along the length of the cavity which would be canceled at the downstream edge corresponds to the idea of a standing wave with a node at the hot-wire location.
However, with a Kelvin-Helmholtz instability, we are in the case of a convective instability which can only be killed by canceling the disturbances at the source.
Otherwise, any oscillations close to the most dangerous frequency would inevitably be amplified and be present at all $x$ and especially at the hot-wire location.
This is well observed in the visualization of the natural flow where the oscillations reach maximum non-linear amplitudes and break on the downstream corner.
Hence, the downstream stabilization of the flow results necessarily on the control of the mixing layer in the streamwise direction.


%% file: S6.tex
\section{Conclusions}\label{Sec:Conclusions}
This paper deals with a closed-loop stabilization experiment of an oscillating flow 
exhibiting non-linear coupling between several frequencies.
The control law is automatically learned with gradient-enriched machine learning control (gMLC) \citep{CornejoMaceda2021jfm}.
The chosen plant is an open cavity flow in experiment for two distinct regimes: 
a narrow-bandwidth regime with dominant frequency $f_a$ 
and a mode-switching regime where another frequency $f^+$ temporarily occurs.
The flow is actuated upstream at the leading edge with a DBD plasma actuator
and monitored downstream at the trailing edge with a hot-wire sensor.
The cost function penalizes the energy peaks at the dominant frequencies in this velocity signal
and the actuation power.

First, the effect of steady forcing is explored.
For the narrow-bandwidth regime, an increasing actuation level 
progressively mitigates  the main frequency $f_a$ while the energy of the other mode ($f^+$) rises.
The fluctuation energy is reduced by up to $97\%$ compared to the unforced case.
The corresponding maximum actuation level defines the limit where a residual resonance can still be observed.
Similarly, on the mode-switching regime, the two frequencies present in the flow ($f_a$ and $f^+$) 
are both damped as the actuation level increases.
$90\%$ decrease of the maximum power is achieved for $88\%$ of the maximum actuation level.
Thus, reducing the main oscillations of the mixing layer is possible with a high-amplitude steady forcing.

Second, a feedback control law from hot-wire signal to DBD actuation 
is optimized with gradient-enriched machine learning control.
The control law associated with the narrow-bandwidth regime
reduces the energy of the peak frequency  to $1.29\%$ of the unforced case,
i.e., more than the steady forcing.
In addition, this better feedback performance requires less than $1\%$ of the steady open-loop actuation power.
Feedback is demonstrated to be crucial for the established control:
an open-loop control with the recorded feedback actuation command has hardly any stabilizing effect.
A novel cluster-based investigation of the control law 
indicates a similar mechanism as fixed point stabilization with phasor control.
This mechanism is corroborated by an analytical simplification of the control law.
Intriguingly,  the phase delay strongly varies with amplitude of the oscillations.
Thus, the control has the features of stabilizing control of fixed point 
with minimal actuation power to compensate for system noise.

Third, gMLC is also employed to optimize the control law 
to stabilize the mode-switching regime.
The learned law manages to successfully decrease the energy related 
to the two main frequencies to $3.35\%$ of the unforced case and 
also with small actuation power around $2\%$ of the maximum actuation level.
This time, the control performs in open-loop as well as in closed-loop.
The actuation mechanism seems hardly interpretable and more complex than phasor control.
Re-evaluation of the learned laws leads to a slight performance drop
rendering them insensitive to the varying experimental conditions.

\begin{figure}
\centerline{\includegraphics[width=\linewidth]{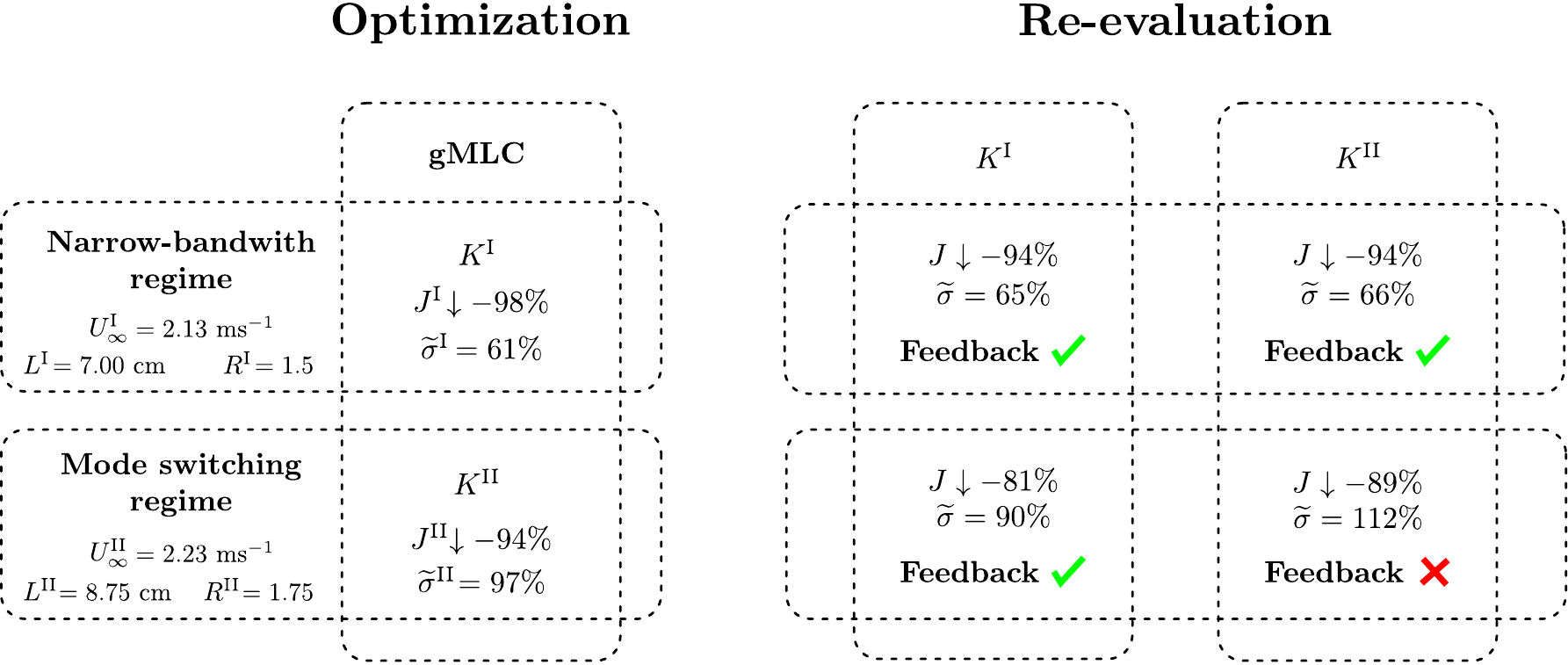}}
\caption{Summary of the performance for the laws learned ($\KgMLCRa$, $\KgMLCRb$) in this study.
On the left, the performance of the laws during the optimization process and on the right, the performance during the offline re-evaluations.
The re-evaluation results are averaged over 20 realizations.
The results associated with the learning on the narrow-bandwidth regime and the mode-switching regime are marked by the superscripts $\textrm{I}$ and $\textrm{II}$ respectively.
The downward arrows symbolize cost reduction.
The feedback symbols indicate whether (\checkmark) or not ($\times$) feedback is a necessary feature for the control:
$\widetilde\sigma$ designates the normalized standard deviation of the downstream velocity.}
\label{fig:ControlSummary}
\end{figure}

Finally, the robustness of the optimized controllers is assessed
by  applying the law learned for one regime  to the other regime.
Expectedly, the law learned in the narrow-bandwidth was only  partially stabilized the mode-switching regime:
The energy associated with frequency $f_a$ is similar to their minimal actuated level
while the energy of $f^+$ is hardly mitigated.
On the other hand, the law learned in the mode-switching regime performs even better than the law learned in the given regime.
The main frequency $f_a$ is controlled and the $f^+$ spillover effect is prevented, 
revealing that a simultaneous control of both frequencies is possible.
Moreover, the need for feedback is demonstrated: 
Applying the recorded closed-loop actuation command
in open loop fashion has hardly any stabilizing effect.
Figure~\ref{fig:ControlSummary} summarizes the control performances for each case and also re-evaluation tests of the learned laws to assess their robustness.
Lastly, the global nature of the stabilization is discussed.


Summarizing, the feedback in stabilization is demonstrated as for similar linear control \cite{Rowley2006arfm} and model-based control \cite{Samimy2007jfm2}.
The actuation power is shown to be a tiny fraction as compared to stabilizing steady actuation.

The key enabler for the fast learning of feedback control laws directly in the plant is 
gradient enriched machine learning control as regression solver.
Genetic programming as evolutionary algorithm explores and populates new local minima
while the subplex simplex method efficiently slides down towards the minima exploiting the local gradient information.
A comparison between gMLC and MLC confirms the benefits of the gradient-augmented method 
for the control performance and learning rate.
Fast learning is critical for experiments with limited testing budget.

Moreover, the performances of the learned laws in one regime at least partially persist when applied to another regime.
Intriguingly, 
the law obtained in the mode-switching regime outperforms the feedback law for single-frequency regime
as it has learned to stabilize the two characteristic frequencies.
\citet{Parezanovic2016jfm} made a similar observation for the destabilization of the mixing layer.
 
We demonstrated the learning capability of gMLC for moderate Reynolds numbers on a single-input single-output (SISO) control experiment.
Ongoing work focuses on the learning of multiple-input multiple-output (MIMO) feedback laws in more complex flows.
One example is drag reduction of a generic truck model under yaw.
Another example is lift increase of an airfoil under angle of attack at a Reynolds number near one million.
Hitherto, already the gMLC predecessor,  machine learning control (MLC)
has been successfully employed in dozens of numerical and experimental plants \citep{Ren2020jh, Noack2019springer} 
comprising $O(10)$ control inputs and $O(10)$ control outputs.
Future, MIMO control law optimizers may be expected to synergize a spectrum of methods.
One example is  cluster-based control  \citep{Nair2019jfm} which can rapidly learn smooth control laws 
and deep reinforcement learning \citep{Rabault2019jfm,Rabault2020joh,Fan2020pnas,Ren2021pof} which seems to be very efficient 
in exploiting short-term actuation responses.


%% file: Acknowledgements.tex
\section*{Acknowledgements}
This work is supported by the French National Research Agency (ANR)
via FLOwCON project ``Contr\^ole d'\'{e}coulements turbulents en boucle ferm\'ee par apprentissage automatique''
funded by the ANR-17-ASTR-0022,  
the iCODE Institute research project of the IDEX Paris-Saclay and by the Hadamard Mathematics
LabEx (LMH) through the grant number ANR-11-LABX-0056-LMH in the
``Programme des Investissements d'Avenir''. 
BRN acknowledges  support  
by the National Science Foundation of China (NSFC) through grants 12172109 and 12172111,
by the Natural Science and Engineering grant 2022A1515011492 of Guangdong province, P.R. China
and appreciates generous technical and scientific support 
from the HangHua company (Dalian, China).

The authors thank Nan Deng and Luc~R. Pastur
for fruitful discussions and enlightening comments.

\section*{Declaration of interests}
The authors report no conflict of interest.

%% file: SA.tex
\section{Main oscillation modes in the open cavity flow}  \label{appA}
As indicated in \S~\ref{Sec:OpenCavityExperiment}, our data shown in figure~\ref{fig:fromBasley2103} correspond precisely to the more comprehensive results from \citet{Basley2013}, provided that the boundary layer thickness at the cavity entrance is slightly corrected.
\citet{Basley2013} present a space-time and frequency analysis of time-resolved velocity measurements recorded at all points of the open cavity flow in the incompressible limit. 
The authors give the origin, the coupling and the prediction of the main spectral features observed when the aspect ratio of the cavity $L/D$ varies between $R=1$ and $R=2$.
The dynamics studied comprise several frequencies $f_\Delta$, $f_\Omega$, $f_b$, $f_l$, $f_a$, $f_r$, $f^+$ that interact nonlinearly with each other and their harmonics.
The first two are shown as originating from centrifugal instabilities taking place span-wise within the intra-cavity recirculation, $f_b$, the so called edge frequency, and all the following ones are directly associated with the shear layer instability.

\begin{figure}
 \centerline{\includegraphics[width=0.3\linewidth]{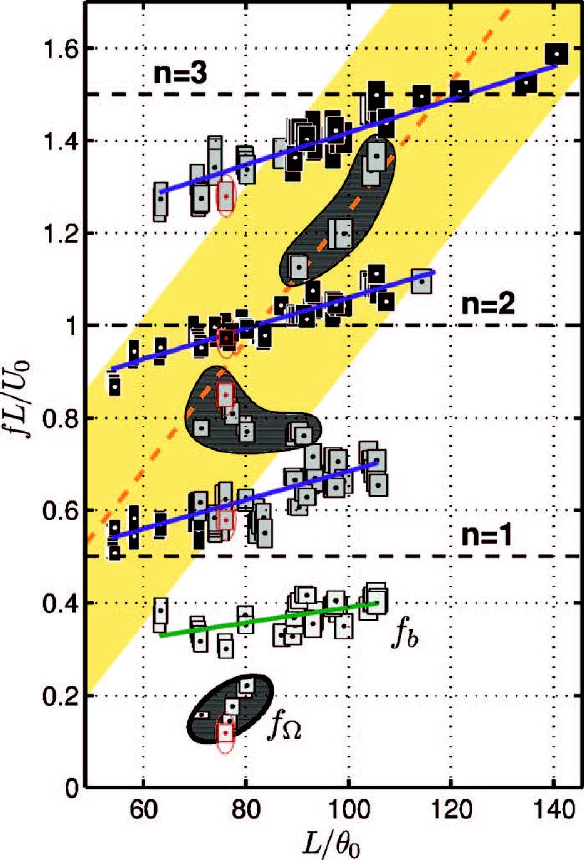}}
 \caption{Strouhal numbers for the main oscillation modes are displayed as functions of the dimensionless cavity length $L/\theta_0$.
Self-sustained oscillations frequencies (black boxes), side-band peaks (gray boxes), and low frequencies (white boxes). 
Rectangle dimensions represent uncertainties.
The shaded area (yellow) is drawn a posteriori such as to segregate self-sustained oscillation frequencies from most side-band peaks.
It is delimited by $St^{L} = St^{L}_{*} \pm 1/3$,with 
$St^{L}_{*} = 0.014 \left( L/\theta_0-10 \right) $ 
the centerline Strouhal number. Hatched regions highlight side-band frequencies departing from the general scheme.
Note that in this graph $U_0$ corresponds to $U_\infty$ of the present paper.
Figure and legend reproduced from \citet[Fig. 3]{Basley2013}, with the permission of AIP Publishing.
}
\label{fig:BasleyFig3}
\end{figure}

We shall not describe the dynamics of the flow as it is presented in detail in \citep{Basley2013}.
Figure~\ref{fig:BasleyFig3}, extracted from \citep{Basley2013}, highlights the interest of the open cavity as a benchmark of adjustable complexity for the development of machine learning algorithms.
Beyond the benchmark role, the open cavity is still one of the flow configurations frequently encountered in industrial applications such as transportation systems and still has a strong impact on the performance and noise level of these vehicles.

%% file: SB.tex
\section{Comparison between gMLC and MLC}  \label{appB}
In this appendix, we compare the learning performance of machine learning control \citep[MLC]{Duriez2017book} and gradient-enriched machine learning control (gMLC, \S~\ref{Sec:gMLC}) for the same experimental conditions as those of \S~\ref{sec:gmlc_R1}, i.e. we aim again to mitigate the oscillations of the mixing layer in the narrow-bandwidth regime.
As a reminder, the narrow-bandwidth regime is described in \S~\ref{Sec:Unforced_dynamics}.

We recall that MLC differs from gMLC in two respects.
First, the evolution consists of improving a group of individuals through generations.
Second, unlike gMLC, no gradient information is employed.
The first generation of randomly generated individuals evolves through the evolution phases thanks to three genetic operations:
\begin{itemize}
\item \emph{Crossover}: two new individuals are generated by stochastic recombination of two individuals, exploiting parts of the \emph{parent} individuals;
\item \emph{Mutation}: a new individual is generated by a stochastic modification in one individual, the resulting individual may share new structures or generate new ones depending on the impact of the change;
\item \emph{Replication}: an identical copy of one individual is generated, assuring memory of good individuals throughout the generations.
\end{itemize} 
The genetic operators are applied to the better-performing individuals to generate the next generations of individuals.
The best individuals are selected with a tournament selection method.
As suggested in \citet{Duriez2017book} a tournament selection of size of 7 for 100 individuals is chosen.
Genetic operations are chosen randomly following given probabilities: the crossover probability $P_c$, the mutation probability $P_m$ and the replication probability $P_r$.
The probabilities add up to unity $P_c + P_m + P_r = 1$.
Following \citet{CornejoMaceda2021jfm}, we choose $[P_c, P_m, P_r ] = [0.6, 0.3, 0.1]$ as this set of parameters converges towards better solutions in average and has one of the lowest dispersion of the final solution.
Moreover, an elitism operator, transferring the best individual of one generation to the next, is employed to assure that the best individual does not get \emph{lost}.
The parameters employed for the definition of the control laws are the same as for gMLC, see table~\ref{tab:cavity_parameters}.
The individuals are evaluated over $T_{\rm ev}=\SI{40}{\second}$ for both the gMLC and MLC experiments.
And for a fair comparison, a population of 100 individuals is chosen to evolve over 10 generations, for a total of 1000 individuals.

\begin{figure}
 \centerline{\includegraphics[width=0.90\linewidth]{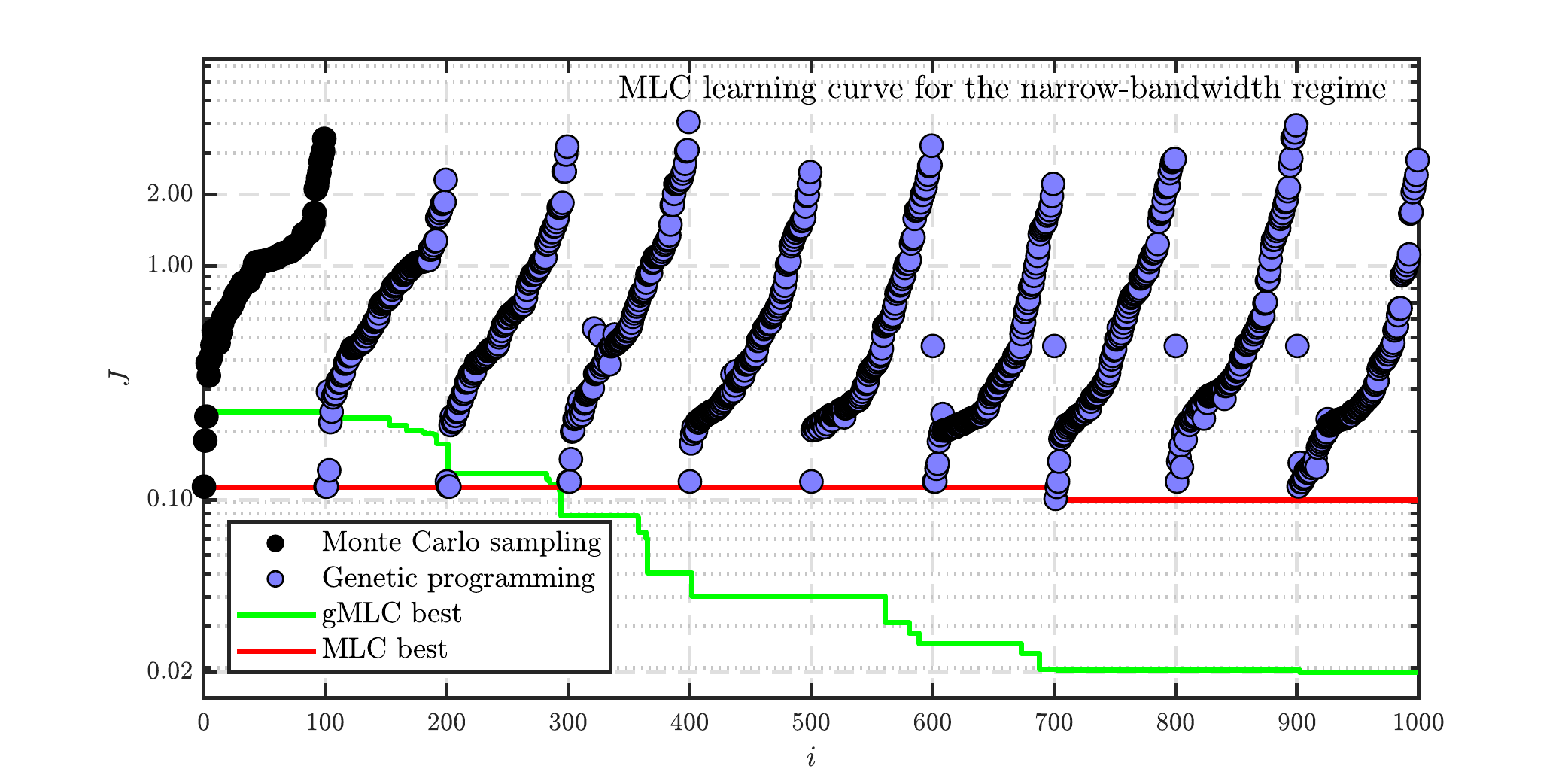}}
 \caption{Distribution of the costs during the MLC optimization process.
 Each dot represents the cost $J$ of one individual.
 The color of the dots represent how the individuals have been generated.
 Black dots for the individuals randomly generated by Monte Carlo sampling (individuals $i=1,\ldots,100$), 
 blue dots for the individuals generated with genetic operator (individuals $i=101,\ldots,1000$).
 For each generation the individuals have been sorted according to their cost.
 The red line shows the evolution of the best cost for the MLC optimization process.
 The green line corresponds to the gMLC optimization process.
The vertical axis is in $\log_{10}$ scale.}
\label{fig:MLCvsgMLC}
\end{figure}

Figure~\ref{fig:MLCvsgMLC} shows the learning process of MLC and the distribution of the individuals evaluated following their cost $J$.
We note that most of the learning is unusually done at the Monte Carlo sampling phase, where the cost is reduced to $J = 0.12$.
The next improvement is carried out at the 8-th generation, where the cost of the best control law slightly decreases to $J=0.10$ and the associated standard deviation is $\widetilde \sigma=1.59$.
Such type of control laws have been encountered in most of MLC realizations.
We take a particular case where the Monte Carlo sampling phase is particularly efficient and where the evolutionary phases does not allow us to leave this local minimum.
It is then necessary to wait for 2000 evaluations to reach performances similar to gMLC, the final cost being $J=0.05$ and the standard deviation dropping to $\widetilde \sigma = 0.73$.
Note that for 700 evaluations, gMLC already reduced the cost function to $J=0.02$.
As described in \S~\ref{sec:gmlc_R1}, the progress of gMLC results on one side, from the exploration of the control law space with the crossover and mutation operators and, on the other side, from the exploitation with the gradient descent performed with downhill simplex.
Therefore, for a same number of evaluations gMLC surpasses MLC both in terms of learning speed and performance of the final solution.
By multiplying the gains in terms of speed and cost, gMLC outperforms MLC by one order of magnitude.
The benefits of gMLC over MLC have been described in \citet{CornejoMaceda2021jfm} for DNS and it is now demonstrated in experimental conditions for the open cavity.

%


%% file: SC.tex
\section{Gradient-enriched MLC laws}  \label{appC}
In this appendix, we describe the control laws derived with gMLC.
In \S~\ref{Sec:K1_description}, we detail the control law learned in the narrow-bandwidth regime $\KgMLCRa$ (\S~\ref{sec:gmlc_R1}) and in \S~\ref{Sec:K2_description}, we give more insight on the control law learned in the mode-switching regime $\KgMLCRb$ (\S~\ref{sec:gmlc_R2}).

\subsection{Control law learned in the narrow-bandwidth regime, $\KgMLCRa$}\label{Sec:K1_description}
The control law $\KgMLCRa$ learned by gMLC in the narrow-bandwidth regime is a linear combination of 19 control laws; it is rewritten in equation~\eqref{eq:K1}.
We recall that the division and logarithm operations are protected to be defined over all the real values.
 We note that $\KgMLCRa$ includes all feedback signals at least one time except sensor $a_8 = u(t-7T_s)$ that is missing.
Sensor $a_1$ and $a_2$ are present in majority; 12 occurrences for $a_1$ and 5 for $a_2$, supporting the possibility of phasor control as a control mechanism.
Table~\ref{tab:control_laws_R1} breaks down the control law $\KgMLCRa$ into linear combination of control laws.
We note that nine additional control laws ($\#11$ to $\#19$) have been introduced in the simplex due to the exploration phases.
Moreover, the best performing control law ($\#15$) among the 19 control laws is associated with the highest weight.
However, control law $\#17$ weight is a close second, suggesting that $\KgMLCRa$ is mainly composed of control laws $\#15$ and $\#17$.
Also, we note that among the five best control laws, they all include mostly sensor $a_1$, advocating that this phase relation may be related to the cavity resonance

\begin{equation}
\label{eq:K1}
\begin{array}{rll}
\KgMLCRa (\bm a) &=&
-0.002318
-0.005459 a_7
-\cfrac{0.000626}{\sin(a_3)}
-0.001108\log(\sin(\exp(a_4)))\\
& & 
-\cfrac{0.009794}{\sin \left( \cfrac{a_2 - 0.079456}{3.2502 \slash a_1} \right)}
+\cfrac{0.001799}{a_5}
-\cfrac{0.002659}{\log(a_6 +2.9498)}\\
& &
+\cfrac{0.023733}{\sin \left( \cfrac{0.668879}{3.2502 \slash a_1} \right)}
+0.15732\sin(3.2502 + a_1)
+0.053174\sin(\log(a_4 - a_2))\\
& &
-0.031016\tanh( \log(a_{10}) - \sin(a_2)  (2.9498-1.00278  a_1))\\
& &
+0.46563\sin( 2.71881 + a_1 + \sin(\tanh(\sin(a_2))))\\
& & +0.08288\tanh(a_1-3.48119 + \sin(\tanh(\sin(a_2))))
+0.43035\sin(3.190735 + a_1)\\
& &
-0.16553 \left( \log(a_1) - \exp \left( 1.526921 \cfrac{a_9 \slash \exp(a_1)}{a_9 + \exp(a_1)} \right) \right)\\
& &
+0.048964\sin(2.71881 + a_1 + \sin(\tanh(\sin(a_1)))),\\
\JagMLCRa &=& 0.0129,\\
\JbgMLCRa &=& 0.0063,\\
\siggMLCRa &=& 60.59 \%.
\end{array}
 \end{equation}
 
\begin{table}
  \begin{center}
\def~{\hphantom{0}}
\sisetup{scientific-notation = true}
\begin{math}
\begin{array}{cccc}
\# & b & \text{Weight} & J\\[3pt]
\midrule
1 & -0.486443 & \num{0.0016} & 0.2783 \\
2 & -0.14431 a_7 & \num{0.0378} & 0.2888 \\
3 & -0.190619 & \num{-0.0040} & 0.3093 \\ 
4 & \cfrac{-0.14431}{\sin(a_3)} & \num{0.0043} & 0.2410 \\
5 & -0.230847 & \num{0.0059} & 0.2770 \\ 
6 & -0.293101& \num{0.0032} & 0.2803 \\ 
7 & \log(\sin(\exp(a_4))) & \num{-0.0011} & 0.2796 \\
8 & \frac{-0.14431}{\sin\left(\frac{a_2 - 0.079456}{3.2502 \slash a_1}\right)} & \num{0.0679} & 0.2276 \\ 
9 & \frac{-0.14431}{a_5} & \num{-0.0125} & 0.3073\\ 
10 & \frac{-0.14431}{\log(a_6 +2.9498)} & \num{0.0184} & 0.2589 \\ 
11 & \frac{-0.14431}{\sin\left(\frac{0.668879}{3.2502 \slash a_1}\right)}  & \num{-0.1645} & 0.1329 \\ 
12 & \sin(3.2502 + a_1) & \num{0.1573} & 0.0967 \\ 
13 & \sin(\log(a_4 - a_2)) & \num{0.0532} & 0.0826 \\ 
14 & 1.03805 \tanh(\log(a_{10}) - (\sin(a_2) (2.9498-1.00278 a_1))) & \num{-0.0299} & 0.2344\\
\textbf{15} & \bm{\sin(2.71881 + a_1 + \sin(\tanh(\sin(a_2))))} & \bm{4.656 \times 10^{-1}} & \bm{0.0311} \\ 
16 & \tanh(a_1-3.48119 + \sin(\tanh(\sin(a_2)))) & \num{0.0829} & 0.0492 \\ 
17 & \sin(3.190735 + a_1) & \num{0.4304} &0.0495 \\ 
\textbf{18} & \bm{\log(a_1) - \exp (1.526921 \frac{a_9 \slash \exp(a_1)}{a_9 + \exp(a_1)})} & \bm{-1.655 \times 10^{-1}} & \bm{0.0462}  \\
\textbf{19} & \bm{\sin(2.71881 + a_1 + \sin(\tanh(\sin(a_1))))} & \bm{4.90 \times 10^{-2}} & \bm{0.0360} \\ 
- &\KgMLCRa & - & 0.0192\\
    \end{array}
    \end{math}
    \caption{\label{tab:control_laws_R1}
    Summary of the 19 control laws composing $\KgMLCRa$  described in equation \eqref{eq:K1}.
    For each control law, we present the mathematical expression $b$, its weight in $\KgMLCRa$ and cost $J$.
    The three best performing control laws $\#15$, $\#18$ and $\#19$ are highlighted in bold.
    }
    \end{center}
    \end{table}

\subsection{Control law learned in the mode-switching regime, $\KgMLCRb$}\label{Sec:K2_description}
The control law $\KgMLCRb$ learned by gMLC in the mode-switching regime is a linear combination of several control laws; it is rewritten in equation~\eqref{eq:K2}.
\begin{equation}
\label{eq:K2}
\begin{array}{rll}
\KgMLCRb (\bm a)&=&
-0.045324
+0.10642\tanh(\sin(\log(\log(a_4))))
-0.065719\log(\sin(\exp(a_4)))\\
& &
+0.80925\log((a_1 - a_4))\\
& &
 +0.29696\log(\sin(- \exp(a_3) - a_9 + a_4 -0.022658))\\
& &
-0.00047578\log(\log(\sin(- \exp(a_3) - a_9 + a_4 -0.022658)))\\
& &
-0.092056\log(\sin(- \exp(a_3)  - 0.387155))\\
& &
+0.14116\log(\sin(- a_3-1.022513))\\
& &
-0.23223\log(a_4 \tanh\left(\cfrac{\log(a_4)}{a_6}\right))
-0.19296(a_9 - a_4 -0.63772 )\\
& &
-0.0023331\log(\sin(\sin(-  4.7665\log\left(\cfrac{a_4}{0.424163}\right)))),\\
\JagMLCRb &=& 0.0335,\\
\JbgMLCRb &=& 0.0229,\\
\siggMLCRb &=& 97.37 \%.
\end{array}
 \end{equation}
Interestingly, contrary to the narrow-bandwidth regime, the best control law, $\KgMLCRb$, includes mostly delayed sensor signals: four instances of $a_3$ and nine instances of $a_4$. 
Table~\ref{tab:K_gMLC_R2} details the 11 control laws that constitute $\KgMLCRb$.
We note that the best performing control law ($\#4$) is the one with the highest weight.
It is worth noting that the control law $\#4$ is a function of a phase difference which may be reminiscent of a Pyragas type control \citep{Pyragas1995}.
%
\begin{table}
  \begin{center}
\def~{\hphantom{0}}
\sisetup{scientific-notation = true}
\begin{math}
\begin{array}{cccc}
\# & b & \text{Weight} & J\\[3pt]
\midrule
1 & \tanh(\sin(\log(\log(a_4)))) & \num{0.1064} & 0.2306 \\
2 & -0.19537& \num{0.2320} & 0.2423 \\
3 & \log(\sin(\exp(a_4))) & \num{-0.0657} & 0.1858 \\ 
\textbf{4} &\bm{\log((a_1 - a_4))} & \bm{8.092 \times 10^{-1}} & \bm{0.0713} \\
5 & \log(\sin(- \exp(a_3) - a_9 + a_4 -0.022658)) & \num{ 0.2970} & 0.2332 \\ 
6 & \log(\log(\sin(- \exp(a_3) - a_9 + a_4 -0.022658))) & \num{-0.0005} & 0.1438 \\ 
7 & \log(\sin(- \exp(a_3) - 0.387155)) & \num{-0.0921} & 0.2263 \\
\textbf{8} & \bm{\log(\sin(- a_3 -1.022513)))} & \bm{1.412 \times 10^{-1}} & \bm{0.1014}  \\ 
9 & \log(a_4 \tanh(\frac{\log(a_4)}{a_6})) & \num{-0.2322} & 0.1880\\ 
10 & a_9 - a_4 -0.63772 & \num{-0.1930} & 0.2341 \\ 
\textbf{11} & \bm{\log(\sin(\sin(-  4.7665\log(\frac{a_4}{0.424163}))))}  & \bm{-2.3 \times 10^{-3}} &\bm{0.0960} \\ 
- &\KgMLCRb & - & 0.0564\\
    \end{array}
    \end{math}
    \caption{\label{tab:K_gMLC_R2}Summary of the 11 control laws composing $\KgMLCRb$ described in equation \eqref{eq:K2}.
    For each control law, we present the mathematical expression $b$, its weight in $\KgMLCRb$ and cost $J$.
    The three best control laws $\#4$, $\#8$ and $\#11$ are highlighted in bold.}
    \end{center}
    \end{table}

%% file: SD.tex
\section{Actuation spectral analysis}\label{appD}
In this appendix, we provide the power spectra of the actuation commands for the narrow-bandwidth regime and the mode-switching regime controlled by the control laws $\KgMLCRa$ and $\KgMLCRb$.

\begin{figure}
\centering
\begin{subfigure}{0.47\textwidth}
  \centering
\includegraphics[width=\linewidth]{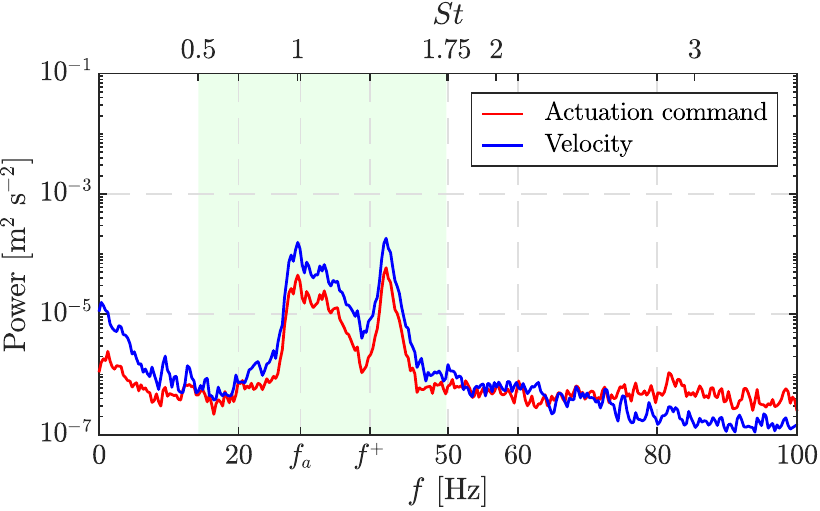}
\caption{
Narrow-bandwidth regime controlled by $\KgMLCRa$.
}
\label{fig:R1K1}   
\end{subfigure}
\hfil	
\begin{subfigure}{0.47\textwidth}
  \centering
\includegraphics[width=\linewidth]{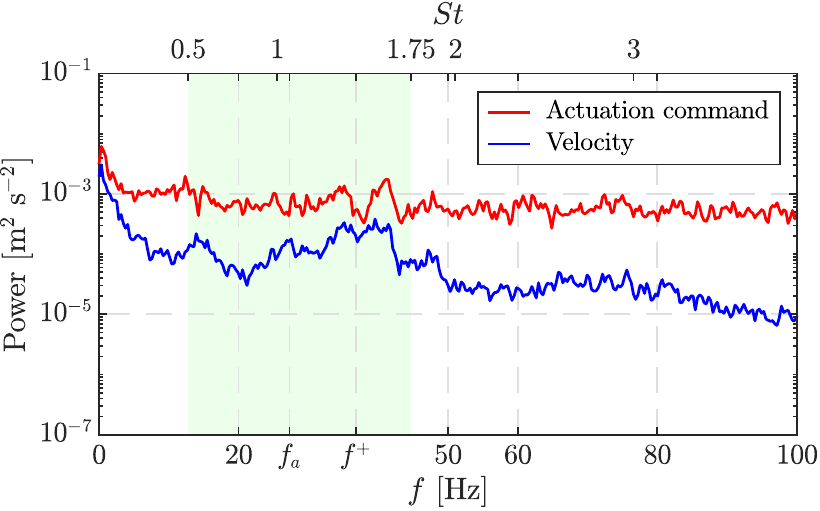}
\caption{
Mode-switching regime controlled by $\KgMLCRb$.
}
\label{fig:R2K2}   
\end{subfigure}

\begin{subfigure}{0.47\textwidth}
  \centering
\includegraphics[width=\linewidth]{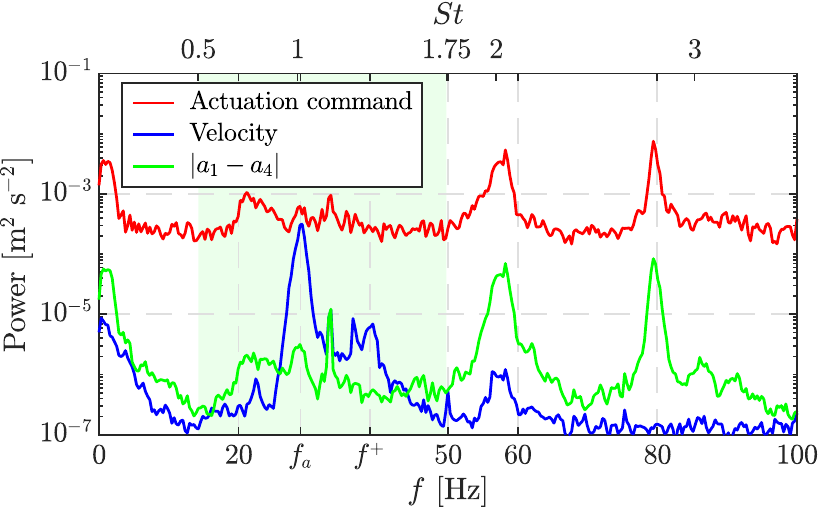}
\caption{
Narrow-bandwidth regime controlled by $\KgMLCRb$.
}
\label{fig:R1K2}   
\end{subfigure}
\hfil	
\begin{subfigure}{0.47\textwidth}
  \centering
\includegraphics[width=\linewidth]{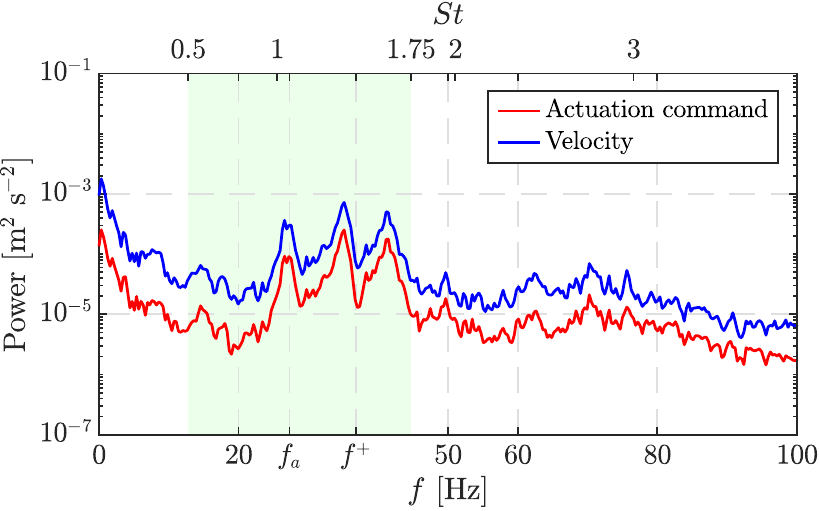}
\caption{
Mode-switching regime controlled by $\KgMLCRa$.
}
\label{fig:R2K1}   
\end{subfigure}

\caption{\label{fig:ControlOtherRegime}
Spectra of the actuation command and the velocity measured by the hot-wire sensor downstream.
}
\end{figure}

When both regimes are controlled with $K^I$ (figure~\ref{fig:R1K1} and~\ref{fig:R2K1}), the actuation command and the velocity measured downstream have a similar spectral signature suggesting a linear relationship between them.
This analysis is partially consistent with the analytical interpretation performed in \S~\ref{sec:gmlc_R1} and \S~\ref{Sec:InterpCross} as $\KgMLCRa$ enables a complex phase relationship between the actuation and the sensing.
%
%

As for $\KgMLCRb$, in both regimes, the spectrum of the actuation command differs from the one of the velocity measured downstream.
For the control of the mode-switching regime (figure~\ref{fig:R2K2}), the spectrum does not display any significant peak except for the very low frequencies.
The actuation command seems to correspond to a random noise without any correlation with the velocity measured downstream, suggesting that $\KgMLCRb$ does not exploit the sensor information for the control, as shown in \S~\ref{sec:gmlc_R2}.
Nevertheless, the control of the narrow-bandwidth regime shows otherwise since the spectrum associated to the actuation command clearly shows two peaks, one at $2f_a$ and $2f^+$ (see figure~\ref{fig:R1K2}).
The presence of the peak at $2f_a$ is consistent with the interpretation of the cluster-based control visualization, suggesting a control at twice the main frequency of the flow $f_a$ (see figure~\ref{Fig:GraphInterpretation_1}).
A systematic evaluation of all the control laws composing $\KgMLCRb$ (see table~\ref{tab:K_gMLC_R2}) shows that it is the term $\log((a_1-a_4))$ that is responsible of the frequency doubling, due to the absolute value function included for the generalization of $\log$ to negative values.
The spectrum of $|a_1-a_4|$ shows the exact same peaks of the actuation command produced by $\KgMLCRb$.
